# Female anatomies disguise ECG abnormalities following myocardial infarction: an AI-enabled modelling and simulation study

Authors: Hannah Smith[1], Abhirup Banerjee[2], Leto Luana Riebel[1], Ruben Doste[1], Maxx Holmes[1], Vicente Grau[2], Blanca Rodriguez*[1]

[1]Department of Computer Science, University of Oxford, Oxford OX1 3QD, United Kingdom

[2]Institute of Biomedical Engineering, Department of Engineering Science, University of Oxford, Oxford OX3 7DQ, United Kingdom

***Address for correspondence.**

Blanca Rodriguez, [1]Department of Computer Science, University of Oxford, Oxford OX1 3QD, United Kingdom, blanca.rodriguez@cs.ox.ac.uk

## Key points summary

- The ECG, important in diagnosing and managing treatment for myocardial infarction (heart attacks), is known to be impacted by the anatomy of the heart and torso.
- Women suffering myocardial infarction are underdiagnosed, and the effect of anatomical sex differences on the ECG in myocardial infarction is poorly understood.
- Human computational modelling and simulation demonstrates that ST-elevation in ischemia was reduced in female anatomies, mainly as their hearts are higher in the torso than in male.
- ECG time intervals were affected differently by anatomy, as T-peak-to-end interval was more sensitive to cardiac positioning, whereas QRS duration to cardiac size.
- Our findings facilitate personalization of clinical guidelines to improve patient outcomes and illustrate how computational modelling and simulation can be exploited to quantify the role of anatomical differences in clinical biomarkers, which may be replicated in many other areas.

## Abstract

The electrocardiogram (ECG) is modulated by torso-heart anatomy, and this challenges patients' diagnosis and risk stratification. This study aims to quantify how torso-heart anatomical factors affect sex-differences in ECG biomarkers in acute and chronic myocardial infarction (MI). We exploit the perfect control of AI-augmented multiscale modelling and simulation, based on clinical magnetic resonance imaging (MRI) data and ECGs for model construction and validation. A cohort of 1720 torso-ventricular anatomies (50% female) was constructed from MRIs of healthy and post-MI participants in the UK Biobank study. 8600 ECG simulations were performed considering anatomical variability and 3 electrophysiological stages (healthy, acutely ischemic, and infarcted). The effect of cardiac size, position, and orientation on each ECG biomarker was quantified. Female anatomies had larger distances between the infarct and ECG electrodes (relative to cardiac size), and larger angles between the infarct normal and ECG lead axes, both primarily caused by their more superior cardiac position. This reduced ST-elevation and caused low-amplitude late depolarisation and repolarisation tails to be missed, shortening QRS duration (QRSd) and T-peak-to-end interval (TpTe). The position and orientation of the heart impacted TpTe more severely than QRSd. AI-enabled mechanistic modelling and simulation identify smaller ventricles, superior cardiac position, and different ventricular orientation as key anatomical contributors of shorter QRS and T waves, and lower ST-elevation, in female versus male anatomies. This provides a blueprint for quantifying the impact of anatomical sex differences on functional markers and enables future work in tailoring clinical guidelines considering unique patient anatomy to reduce biased outcomes.

# Introduction

Anatomical sex differences have been shown to have critical functional consequences in diverse medical specialties ranging from neurology to nephrology [1, 2]. In cardiology, women are more likely to have poor outcomes following myocardial infarction (MI) [3]; whilst the origins of this disparity are multifactorial, including comorbidity profile differences, they are also more likely to have their MI diagnosis missed and undertreated [4-6].

The electrocardiogram (ECG) is widely used in cardiology, and specifically for MI diagnosis. In the weeks following infarction, electrophysiological and structural damage reduces conduction velocity and increases repolarisation heterogeneity in the heart [7, 8], predisposing patients to arrhythmias and subsequent sudden cardiac death [9, 10]. Clinical guidelines utilise ST-elevation for patient risk stratification and selection of reperfusion therapy [11]. Thresholds that define ST-elevation are 0.2mV in men over 40 years old, 0.25mV in men under 40 years old, and 0.15mV in women of any age, all for leads V2 and V3. However, no other leads have sex specific thresholds, and body mass index (BMI) and other anatomical factors are not considered in any lead [12]. Furthermore, ECG metrics such as QRS and T-peak-to-end (TpTe) prolongation have been proposed to more specifically identify risky arrhythmic substrate, but their success in clinical trials has been mixed [13-15]. Anatomical sex differences are overlooked, with a healthy and prolonged QRS duration (QRSd) regarded as <100ms and >120ms respectively, regardless of sex. Given their smaller hearts and shorter QRS, women are more likely to be incorrectly classified as low risk for sudden cardiac death from their ECG [16, 17].

Women have not only smaller hearts, but their heart is also positioned more superiorly and posteriorly in the torso, with different orientation [18-20], resulting in variable effects on specific ECG biomarkers. For example, the shorter female QRSd has been associated with smaller cardiac size whereas the lower female T wave amplitude with larger distances to the exploring electrodes due to heart-torso positioning [19, 21]. However, the rich anatomical and electrophysiological variability of clinical datasets confounds isolation of specific causal links and fails to provide insight as to their mechanistic origin. This is further complicated by patient differences in the nature, location, and size of electrophysiological remodelling following infarction. Torso tank experiments on an acutely ischemic canine heart have shown that rotations of the heart were sufficient to transition from an ST-elevated to electrocardiographically "silent" ECG, but the effect of anatomical sex differences on ECG parameters in diverse anatomies during ischemia is largely unknown [22].

Computational simulations using AI-enabled multiscale models allow precise control of anatomical variability and electrophysiological characteristics. They have proved successful in replicating and investigating the effect of both acute ischemia and infarction on the ECG [23-25]. However, anatomical variability and sex-differences have been largely overlooked. Integrating information from large datasets such as the UK Biobank (UKB) [26], artificial intelligence (AI)-based image processing [19, 27, 28], clinical data analysis, and mechanistic modelling and simulation [29] allows for efficient simulation studies with large scale populations of anatomical models [30] that may be replicated in diverse applications beyond cardiology.

The aim of our study is to quantify the key torso-ventricular anatomical determinants of sex differences in clinical ECG biomarkers used in MI diagnosis and risk stratification. We hypothesise that post-occlusion, ST-elevation, an amplitude crucial in diagnosing myocardial infarction, will be underestimated in female anatomies due to larger distance between infarct and ECG electrodes due to heart-torso positioning. Furthermore, female ECG intervals will be shortened by smaller cardiac size. We exploit the perfect control of simulations using 1720 torso-ventricular anatomies to isolate and quantify the impact of heart size, location, and orientation on the ECG in ischemia and infarction.

## Methods

**Framework – AI-enabled, imaged-based multiscale electrophysiological modelling and simulation**

Figure 1 illustrates the methodologies developed through this study, to compute ECG simulations in multiscale electrophysiological healthy and MI models with 1720 torso-ventricular anatomies.

[19, 24, 29, 31]

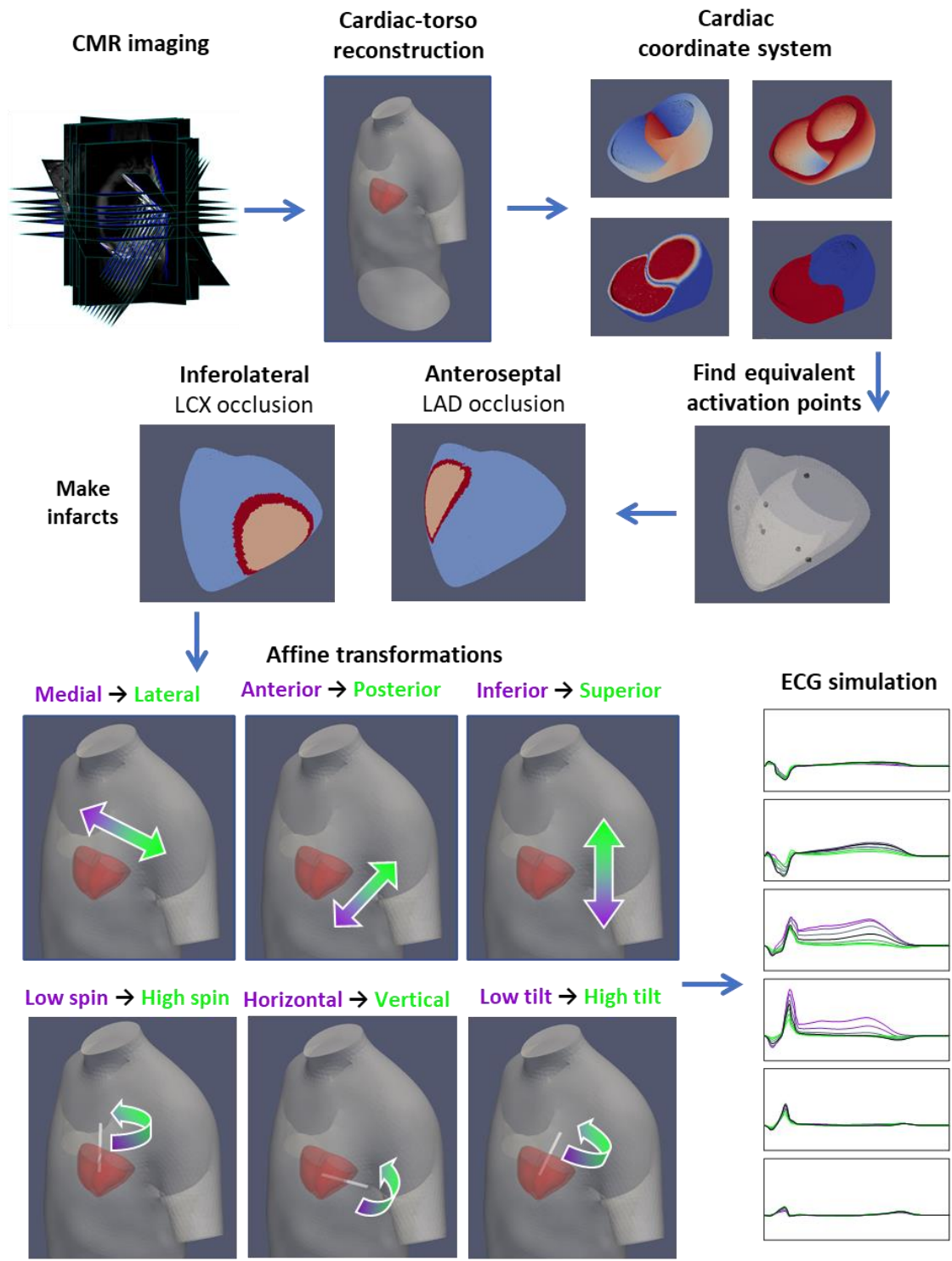

**Figure 1.** *ECG simulations with AI-enabled, image-based anatomical models.* Cardiac magnetic resonance (CMR) imaging was automatically segmented and the resulting contours were used to reconstruct the cardiac-torso anatomy. A universal coordinate system was used to define equivalent activation points and infarct geometries on the cardiac meshes. Electrode positions were altered to mimic cardiac translations in the medial-lateral (x), anterior-posterior (y) and inferior-superior (z) directions, and cardiac rotations to spin

around the torso long axis, in the horizontal-vertical direction, and around the cardiac long axis to tilt the cardiac short axis plane. The ECGs were then simulated using the heart-torso anatomies.

**Subject selection and anatomical reconstruction**

To allow for perfect control of anatomical variations, 1720 anatomies were created by applying 42 affine transformations to the cardiac position and orientation of 40 torso-heart anatomies, which spanned variability in a large population from the UKB [26]. Firstly, 40 subjects were selected from a population of 1476 healthy and chronic post-MI subjects described in Smith et al. 2025 [19]. Ten subjects were selected from each of the healthy male, healthy female, post-MI male, and post-MI female populations to replicate distributions of parameters describing cardiac and torso size, and the relative position and orientation of the heart within the torso. As BMI was a critical determinant of many anatomical characteristics, one subject from each BMI decile was chosen for a population. The final subject selection was manually performed to replicate the range of other anatomical parameters within a population, and the difference in anatomical parameters between populations. The anatomical characteristics of the selected subjects, in comparison with the overall populations are illustrated in Figures 2B-M and Supplementary Table 1.

As illustrated in Figure 1 and previously described in Banerjee et al. 2021 [31], the contours of the myocardial surfaces were extracted using a machine learning network [27, 32], they were aligned to correct for motion artifacts, and then a template geometry was iteratively deformed to fit the aligned contours. The surfaces were then transformed into volumetric tetrahedral meshes using open source MATLAB code from https://github.com/rdoste/InSilicoHeartGen [33], before being converted to hexahedral meshes using open source software from https://github.com/rsachetto/hexa-mesh-from-VTK. The resulting variability in cardiac anatomy, and difference between sexes, is illustrated in Figure 2A.

1720 anatomies were then obtained by applying 42 affine transformations to the electrode positions with respect to the heart. Translations were made in the medial-lateral (x), anterior (y) and inferior-superior (z) directions in 1cm steps by up to ±3cm. Posterior translations of the cardiac mesh were performed in steps of 0.5cm, up to 1.5cm, to prevent the cardiac mesh and electrode positions crossing over. Rotations were made varying the three Euler angles between the torso and cardiac axes. Detailed derivations are given in Supplementary Section 1.2, and the rotations and translations are illustrated in Figure 1. Briefly, the "spin" denoted the angle through which the torso short axis plane was spun around the torso long axis to give the cardiac short axis plane. The verticality of the cardiac long axis in the torso frame was also varied. Finally, "tilt" denoted the angle through which the cardiac short axis plane was tilted by a rotation around the cardiac long axis. All rotations were made in 5° steps up to ±20°. These ranges of parameters used in simulations were chosen to be within the ranges of variation in cardiac position and orientation observed in the healthy and post-MI UKB populations, that are shown in Figures 2E-M.

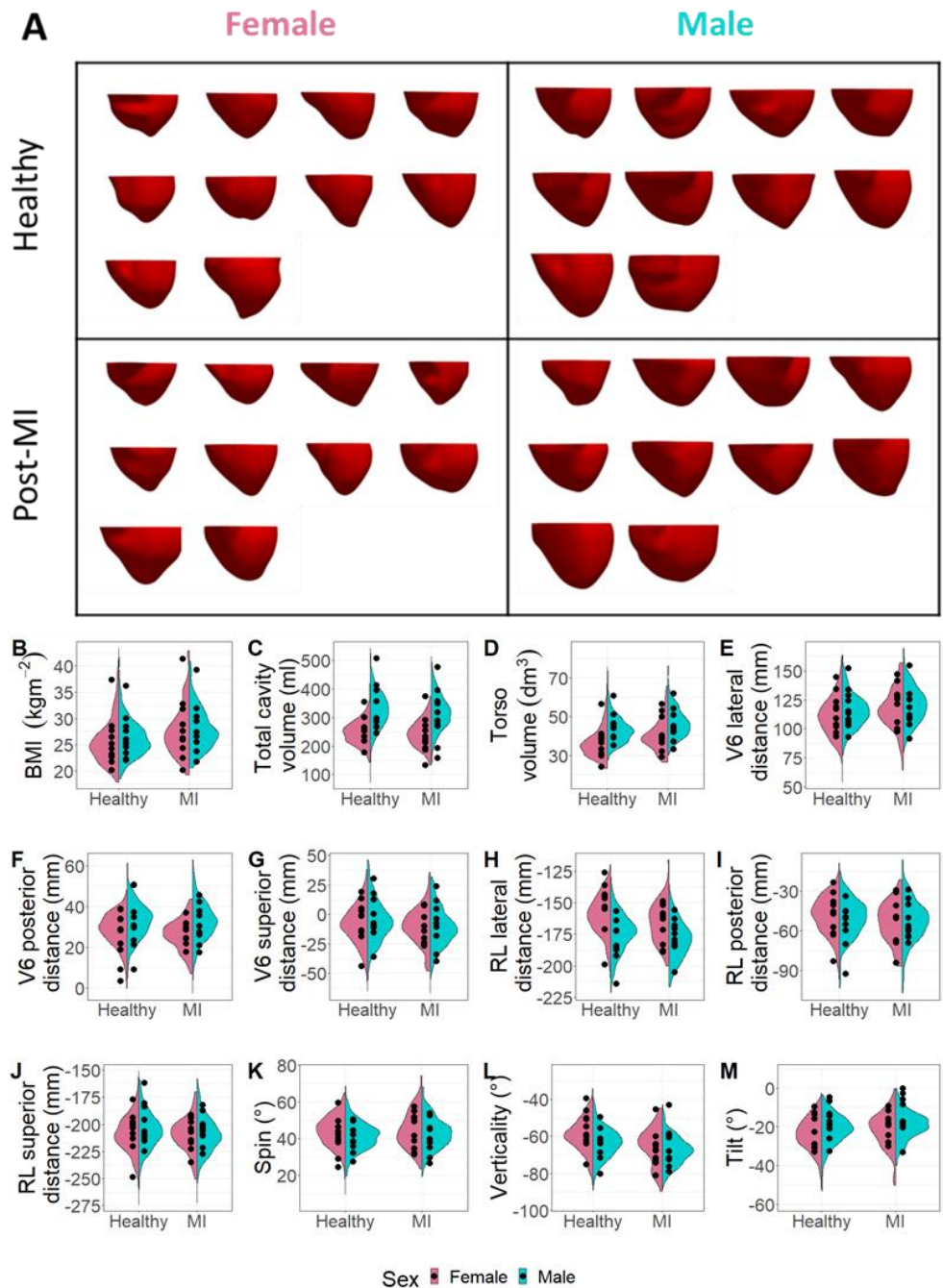

**Figure 2.** *Torso-ventricular size and shape variability*. **A:** reconstructed cardiac meshes arranged by mesh volume in female (left) and male (right) subjects for healthy (upper) and chronic stage post-MI participants (lower), all plotted to same scale. **B-M:** distributions of anatomical parameters in the larger UKB populations are shown in pink (female) and cyan (male) with black points showing subjects selected for detailed reconstruction and simulation. The distances between two electrode locations (**E-G**: V6, **H-J**: right leg, RL) on opposing sides of the torso and the heart centre in three perpendicular directions are shown.

**Multiscale ECG modelling and simulation**

The electrophysiological behaviour of the ventricles was simulated using established modelling and simulation pipelines. Simulations were conducted using the MonoAlg3D simulator, which represents the diffusion of the electrical signal between mesh volumes using the monodomain equation [29]. The human ventricular electrophysiological cell model used was ToR-ORd due to its established credibility in health and disease [34, 35]. For each point, the orientation of the orthonormal conduction directions was assigned in the fibre, sheet, and sheet-normal directions using a rule-based system [36], and the conductivity in each direction was tuned to match both clinically observed human conduction velocities, and clinical values of QRSd [19, 37]. Sinus rhythm was initiated by regularly applying a stimulus current to points on the endocardial surface, establishing a realistic activation sequence that mimics that achieved by the Purkinje network. These early activation points were replicated in equivalent positions across anatomies using a system of universal ventricular coordinates that assigns rotational, transmural, apicobasal, and intraventricular positions to each mesh point (Cobiveco [38]). The activation was efficiently spread by setting the conductivity of the cells in the top 18% of transmurality (i.e. innermost endocardial, tuned to be the smallest proportion that achieved a uniform layer) to a factor of four above normal (tuned to match clinical QRSd [19]). The apicobasal coordinate was used to replicate physiological gradients in IKs by multiplying the conductance of the channel by a factor of $2^{2 \times apicobasal - 1}$, thus from $0.2 \times G_{Ks}$ at the base to $5 \times G_{Ks}$ at the apex, giving a 10ms action potential duration (APD) difference [39]. Volumes were assigned to the correct transmural cell type (endo-, mid- and epicardial) depending on their transmural coordinate, giving a 37ms action potential duration (APD) transmural gradient [40].

**Modelling stages of ischemia and infarction**

Modelling and simulation enables the replication of specific timepoints following coronary occlusion, remodelling characteristics, and infarct sizes and locations, across anatomies [24, 25, 41-43], unlike clinical and experimental approaches. Ionic alterations caused by acutely ischemic (hours) and the healing infarct (weeks) stages were applied to the ventricular myocyte models by changing ionic concentrations, and the conductances and time constants of key ion channels, as in Wang *et al*. 2021 and Riebel *et al.* 2024 [24, 42], with details described in Supplementary Table 2.

Infarcts were placed in anteroseptal and inferolateral locations to model left anterior descending and left circumflex artery occlusions respectively, as shown in Figure 1. For infarction, the affected area was represented as a core infarct zone, with more severely impacted electrophysiological properties (given below), surrounded by a borderzone with properties closer to the healthy values. In order to assign the infarct geometry equivalently across anatomies, ellipsoids in Cobiveco space were defined, and the radii iteratively adjusted until the same proportion of myocardial tissue was affected, matching infarct sizes reported in late gadolinium enhanced cardiac magnetic resonance imaging of post-MI patients [44, 45]. The acutely ischemic zone was taken to be the borderzone and infarct zone combined, to match the larger observed ischemic/infarct sizes in the earlier stages of ischemia/infarction [44, 45]. The infarct/ischemic zones were chosen to be fully transmural.

The electrophysiological characteristics of the ischemia/infarction were as follows. In the acutely ischemic phase, the ischemic region had normal conductivity, but extracellular potassium concentration was increased (8.5mM), the ATP-sensitive potassium current (IKATP) was activated, and the fast sodium (INa) and L-type calcium (ICaL) currents were inhibited, as summarised in Supplementary Table 2. These effects acted to depolarise the resting membrane potential and shorten the APD. In infarction, the extracellular potassium concentration returned to normal (5mM), and IKATP was switched off. However, the IKr, Ito and IK1 potassium currents were reduced, as detailed in Supplementary Table 2. Additionally, ICaL and INa were more strongly inhibited than in ischemia, and calcium kinetics were slowed. This acted to decrease the resting membrane potential back towards normal (-89mV), but the APD was significantly prolonged, particularly in mid-myocardial cells. The borderzone had similar, but less severe remodelling. These electrophysiological changes were based on experimental and clinical data from ischemic/infarcted myocardium [24], and their effect on the cellular action potentials is shown in Supplementary Figure 3. The conductivity in the infarct and borderzone was tuned to half that of normal tissue to fit experimental data on conduction velocity and QRSd, with conduction velocities illustrated in Supplementary Figure 2 [7, 19].

**Simulation protocol**

8600 ECGs were simulated, considering healthy electrophysiology, acute ischemia and infarction in two locations (five total scenarios) over the 1720 anatomies. The cellular electrophysiology model was first ran for 100 beats to achieve a steady state initial condition for the cells in the 3D mesh. Then the individual points were connected to each other via an orthotropic conductivity matrix which allows diffusion of the electrical activation between neighbouring points. The early activation points were stimulated every 1000ms for 3 beats to achieve a heart rate of 60bpm. All simulations were performed using the MonoAlg3D monodomain solver software on the Cirrus supercomputing resource (Edinburgh Parallel Computing Centre, University of Edinburgh), with each cardiac anatomy and infarction combination taking approximately eight hours on one GPU compute node, with an estimated total supercomputing time of 1600 hours.

**Credibility: Verification, validation, and uncertainty quantification**

Consideration was given to verification, validation, and uncertainty quantification of each component of the proposed framework in order to build its credibility [46]. The MonoAlg3D monodomain solver software has previously been verified against other solver platforms using slab-based and whole-ventricle benchmark tests [29], and convergence analysis on conduction velocity has been performed [42]. The morphological characteristics of the acutely ischemic and infarcted simulated ECGs obtained in this work were comparable with those obtained by Wang *et al*. 2021 [24], and in agreement with clinical data across infarct stages [42].

The human ventricular cellular model has repeatedly shown consistency with cellular and clinical data in healthy, drug-block, and disease conditions, including hyperkalaemia [34]. In this work, simulated healthy and infarcted ECG morphology and biomarkers were compared with the UKB clinical population, and the acutely ischemic stage ECGs with clinical ECGs from the Physionet database [47], to support credibility for the simulations in this context.

Sensitivity analyses have been performed to identify the currents primarily responsible for cellular electrophysiological behaviour, and a population of models framework has previously been applied to them in order to probe the effects of interpersonal variability [41, 48]. In this work, the sensitivity of the ECG to the conductivity of the infarct was tested, as illustrated in Supplementary Figure 4.

**ECG calculation and delineation**

Simulated ECGs was calculated in MonoAlg3D using the pseudo ECG method [29, 49]. A custom MATLAB script was developed to delineate the ECG, with further details in Supplementary Section 1.6. Code is available in https://github.com/MaxxHolmes/ECG_Delineation. The method operates on the principle of gradient analysis of sliding windows. QRSd, TpTe, and ST amplitude were selected as biomarkers because of their relevance to the diagnosis of, and risk stratification following, myocardial infarction, and their representation of different sections of the cardiac cycle. All markers were calculated on the final beat.

**Statistical analysis**

The proportion of the sex difference in each ECG parameter that was caused by sex differences in three anatomical parameters was estimated as follows. First, the myocardial volume was calculated as the number of hexahedral elements in the cardiac anatomy multiplied by the volume of each element. This served as a parameter for the overall size scale of the anatomy. The distance from the exploring electrode to the centre of the infarct region was calculated. This was then normalised by the cube root of the cardiac volume, to effectively measure the distance in "cardiac lengths", in order to remove the effect of overall scaling and reduce collinearity. The angle between the ECG lead axis (the vector from the exploring electrode to Wilson's central terminal near the centre of the heart) and the normal to the surface at the infarct centre (extracted using ParaView [50]) was calculated. Linear models were then obtained, pooling all the cardiac and torso anatomies and precordial leads (V1-V6). Variance inflation factor values for each linear regression parameter set are given in Supplementary Table 3, showing a low degree of collinearity, with a maximum value of 1.4. The precordial leads were chosen because they are all unipolar, and the distances between them are small enough that the simplification of assuming linearity is more valid. The mean difference in the anatomical parameter between male and female anatomies was then multiplied by the regression coefficient for that parameter to estimate its contribution to the sex difference in the ECG biomarker. For the acutely ischemic simulations, the number of simulated subjects with diagnosable contiguous ST-elevation, according to existing clinical guidelines [11], was counted both in the unaltered cardiac position, and with each cardiac translation and rotation.

# Results

## Simulated ECGs match clinical patterns throughout stages of ischemia and infarction

Figure 3 illustrates the comparison between simulated ECGs (black, A-D) and clinical ECGs (green, B-D) in healthy, acute ischemic, and infarction conditions, alongside associated activation and repolarisation maps (B-D) and sex-specific distributions compared with UKB data (E). Note that simulated healthy ECGs showed physiological R wave progression and upright T waves across all precordial leads (Figure 3A), and for all disease states, ECG morphology was similar in simulated and clinical ECGs in healthy, ischemic, and infarcted ventricles (Figures 3B-D), with disease-related abnormalities explained by activation and repolarisation maps.

With healthy electrophysiology, the mean QRSd across leads was shorter for female than male geometries at 71.6±6.8ms and 79.7±9.3ms respectively (Figure 3E, left panel). This is consistent with sex differences in the UKB healthy population (mean QRSd of 79.8±13.3ms and 85.8±10.8ms for women and men respectively). Simulated healthy TpTe was also shorter for female versus male geometries at 60.7±13.2ms and 64.2±13.5ms respectively, as reported in Smetana et al. 2003 [51], as shown in Supplementary Figure 5.

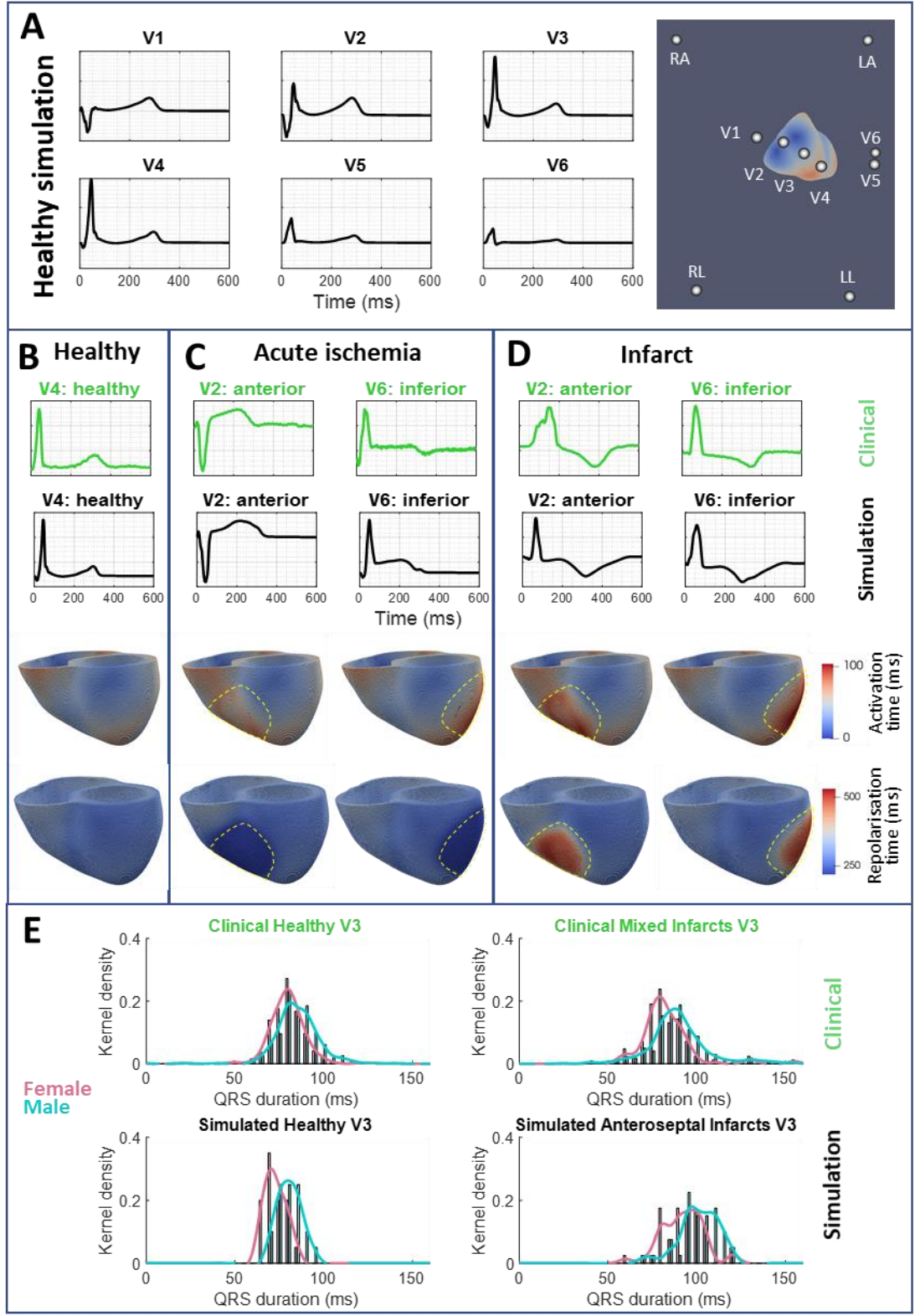

**Figure 3.** *Similarity between simulated and clinical ECGs in all simulated stages of myocardial infarction*. **A**: healthy ECG simulation example in all precordial leads (left) and electrode placement (right). Comparison between clinical ECGs from the UK Biobank and Physionet (upper, green) and simulated ECGs (lower, black) in healthy (**B**), acutely ischemic (**C**), and infarcted (**D**) hearts. Infarcts were placed in anteroseptal and inferolateral

locations. Acute ischemia showed ST-elevation, and infarct showed T wave inversion and QRS and T wave prolongation, in both clinical and simulated ECGs. Activation and repolarisation maps are shown for an example simulation. **E**: QRS duration distributions in UK Biobank subjects (top panels), for healthy (left, 470 male and 581 female) and mixed location post-MI (right, 341 male and 84 female) in V3 versus simulated QRS durations (bottom panels) in healthy electrophysiology (left) and healing stage post infarct (right) in V3. For both simulated and clinical ECGs, QRS was prolonged in infarcted versus healthy subjects and female QRS durations were shorter than male.

Simulated ECGs in acute ischemia were characterised by substantial ST-elevation (Figure 3), albeit smaller in female than in male anatomies (e.g. mean ST amplitude in lead V6 was elevated to 0.41±0.15mV versus 0.49±0.14mV for female and male anatomies in inferolateral ischemia). For anteroseptal ischemia, ST-elevation was most prominent in leads V1-4, and for inferolateral, in leads V5-6. These patterns are consistent with clinical ECGs in the Physionet database [47].

In simulations, infarction led to QRS prolongation, and severely inverted and prolonged T waves, as in some clinical ECGS in post-MI UKB data (Figure 3D). Figure 3E shows that for both simulated and clinical ECGs, healthy and infarcted QRSd were shorter in female than male subjects. This is consistent with a clinical study reporting QRSd in V3 was 84.8±14.6ms and 91.4±17.5ms, in women and men respectively [19]. For simulated transmural infarcts, QRSd in V3 was 93.8±12.3ms and 103.2±11.1ms in female and male anatomies respectively. The consistency of clinical and simulated sex-differences in biomarkers, and the effect of ischemia and infarction, lends credibility to the modelling and simulation investigation.

## Due to their smaller size, only 20% of female anatomies have QRS prolongation post-occlusion versus 55% of male

Figure 4 quantifies the sex differences in QRSd and TpTe at different stages post-occlusions and their main underlying anatomical causes. Figure 4A illustrates that, for all stages and locations, QRSd was up to 11±3% shorter in female anatomies, with a lead-by-lead analysis in Supplementary Figure 5. Thus, 55% of infarcted male anatomies had maximum QRSd prolonged above 100ms, versus only 20% of female anatomies. Figure 4B reveals what proportion of this sex difference was explained by each anatomical factor: cardiac size, distance of ECG exploring electrode to infarct centre, and angle between ECG lead axis and infarct normal. The 28±7% smaller ventricular size in females was the principal anatomical determinant of QRSd, accounting for between 85% and 91% of the total sex difference.

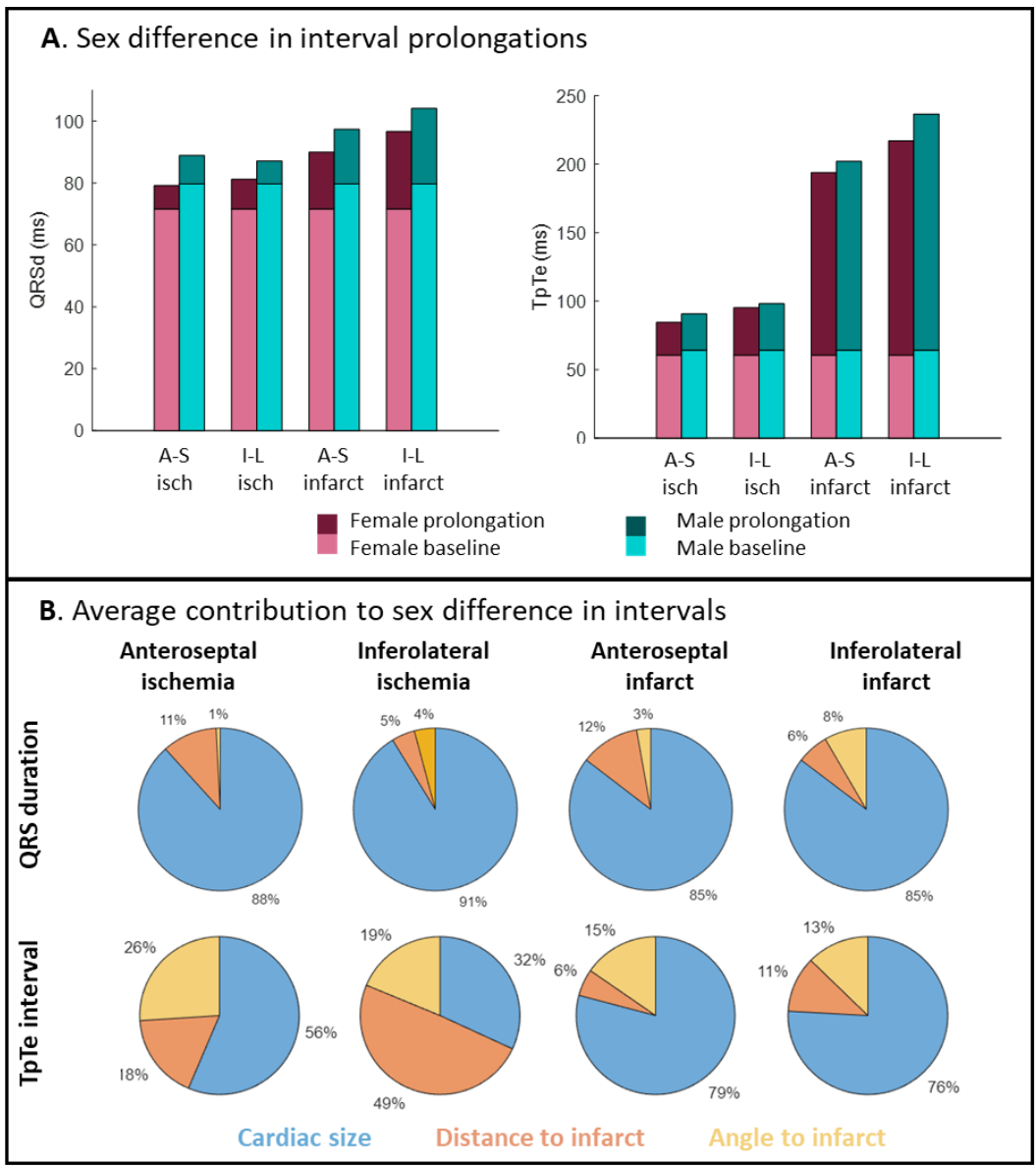

**Figure 4.** *Sex difference in interval prolongations from ischemia and infarct and their principle anatomical contributors.* **A:** Simulated QRS duration (QRSd, left) and T-peak-to-end interval (TpTe, right) in female (pink) and male anatomies (cyan) for healthy electrophysiology (lighter) and prolongations caused by ischemia/infarct (darker). The mean across leads is shown. A-S: anteroseptal, I-L: inferolateral, isch: ischemia. **B:** Estimate of each anatomical parameter's contribution to the sex difference in QRSd (upper) and TpTe (lower), calculated as the product of the correlation coefficient of a parameter and the sex difference in that parameter, averaged over the precordial leads. Yellow: volume of the cardiac myocardium, red: distance from the exploring ECG electrodes to the centre of the ischemic/infarcted region, yellow: angle between the ECG lead axes and the normal to the cardiac surface at the infarct centre.

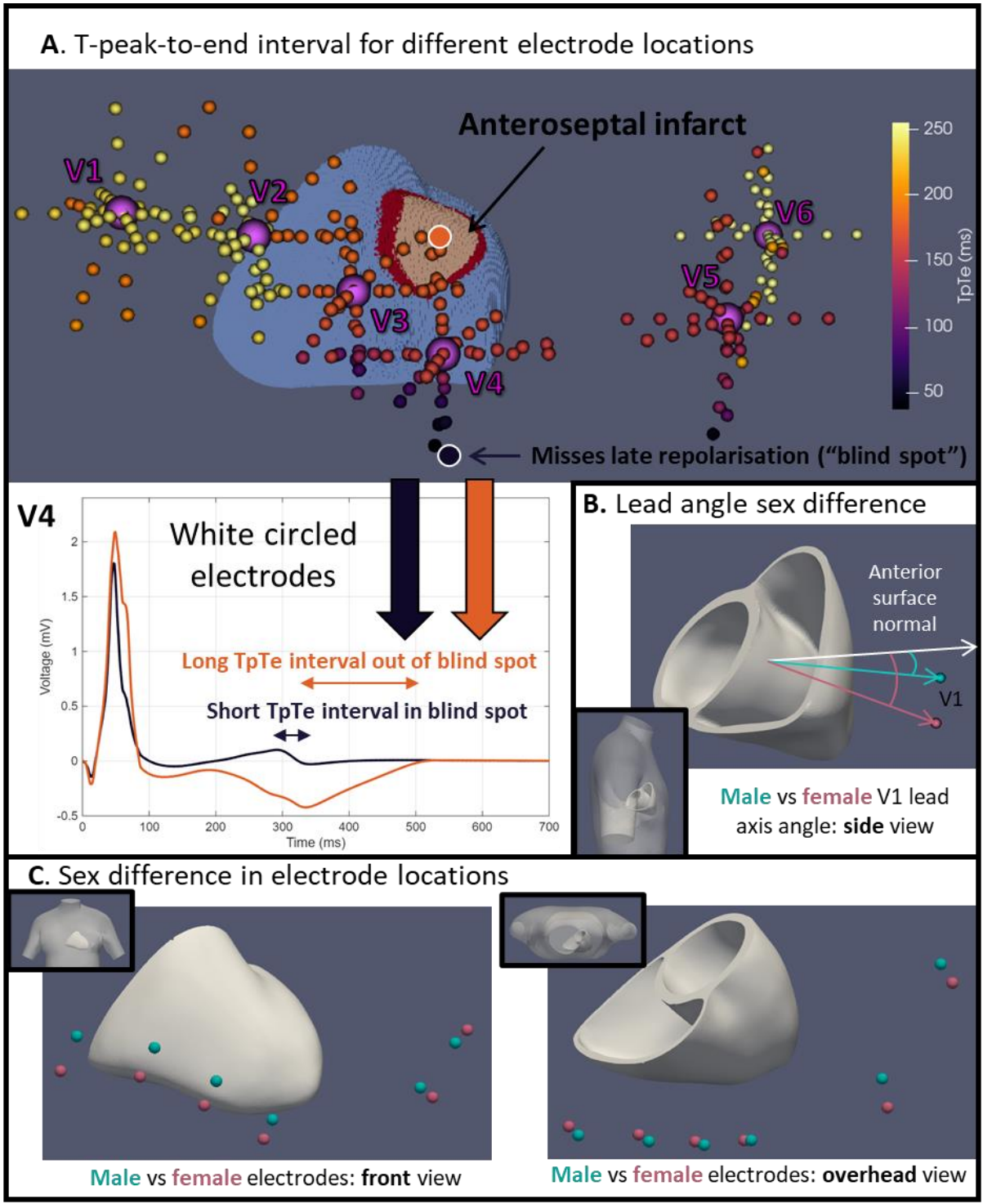

**Figure 5.** *Sex difference in electrode location and effect on T-peak-to-end interval.* **A:** all tested exploring electrode positions (including rotations and translations of the heart) for one example cardiac anatomy. They are coloured by their T-peak-to-end (TpTe) interval in anteroseptal infarct. Purple electrode locations are the original (not translated/rotated) locations and the infarct (beige) and borderzone (red) regions are shown on the cardiac surface. Lower: corresponding V4 ECG trace for electrode locations circled in white (only differing in superior cardiac position) coloured by their TpTe interval. **B:** difference in the angle between the average V1 lead axis angle (male green, female pink) and the normal to the anterior cardiac surface (white) on an example cardiac anatomy from a side on view. **C:** average male (green) and female (pink) electrode position relative to cardiac centre, corrected for cardiac orientation and cardiac size, from a front (left) and overhead (right) view.

## Shorter female TpTe prolongation post-occlusion was caused by larger distances and angles from infarcts to ECG electrodes due to superior heart positioning inside the torso

For both sexes, TpTe interval exhibited prolongation post-occlusion, enhanced in infarction versus ischemia. This was secondary to severe repolarisation delays in the infarct, as illustrated in Figure 3D. As for QRSd, TpTe interval was shorter for female than male anatomies, by up to 20±5ms on average (Figure 4A and Supplementary Figure 5). In this case however, smaller cardiac size only contributed between 32% (inferolateral ischemia) and 79% (anteroseptal infarct) of the sex difference in TpTe interval (Figure 4B). Cardiac size most impacted TpTe in anteroseptal infarction, with every 10ml increase in myocardial volume increasing the TpTe by 6.8±0.2ms, and least in inferolateral ischemia, with every 10ml increase in myocardial volume actually decreasing the TpTe by 1.3±0.1ms.

Sex differences in TpTe interval were partly explained by the angles between the ECG lead axes and the normal to the cardiac surface at the infarct centre. Different positions/orientations of the heart with respect to the electrodes alone (with the same cardiac anatomy) led to a difference in TpTe interval of up to 323ms, equivalent to a ten-fold difference. Figure 5A shows the TpTe interval (colour) for each precordial electrode position in an example cardiac anatomy. It demonstrates that electrode locations that had a large angle between the cardiac surface normal at the infarct centre and the ECG lead axis missed the long repolarisation tail. The electrode locations shaded dark blue were in a "blind spot" that did not record the late repolarisation from the infarct at all – particularly centred on inferior electrode locations around the anterior surface (V3 and V4), where female electrodes were more likely to be. The lower panel of Figure 5A illustrates the ECG recording for the electrodes circled in white, in and out of this blind spot. Females displayed a more superior cardiac position with respect to the electrodes, leading to larger angles by up to 6.8±3.4° (Supplementary Table 5) (Figure 5B).

Figure 5C illustrates the average sex difference in electrode position relative to the cardiac centre, corrected for ventricular size and orientation, with further details in Supplementary Table 4. It demonstrates that the female electrodes were relatively further from the cardiac surface, and thus the exploring electrode was up to 0.33±0.15 cardiac lengths from the centre of the infarct. This increased electrode-infarct distance in female anatomies most substantially affected ischemic rather than infarcted TpTe (anteroseptal: 18% of total sex difference, inferolateral: 49%). For ischemic ECGs, this relationship was driven by the peak T time rather than the end T time. If the ischemic region was near enough the exploring electrode, peak T time moved earlier due to the early repolarising ischemic tissue, elongating TpTe. This increased electrode-infarct distance also shortened QRSd in female anatomies, however this was simply as the low amplitude late depolarisation tail were reduced in amplitude, and thus missed on delineation. Further details for the effect of each cardiac translation and rotation are given in Supplementary Figure 7.

## Cardiac positioning alone changes diagnosis of ST-elevation in 75% of female cardiac anatomies versus 65% of male cardiac anatomies.

Sex differences in ST changes are illustrated in Figures 6A (ischemia) and Supplementary Figure 6 (infarction). They demonstrate that the ST-elevation in acute ischemia, and depression in infarction, were markedly smaller in female versus male anatomies. Lead V4 in anteroseptal ischemia was particularly affected, with mean ST-elevation to 0.185±0.131mV for female anatomies versus 0.332±0.355mV for male. This is in contrast with clinical thresholds for diagnosing ST-elevation being the same for men and women in this lead (0.1mV for both).

Figure 6B illustrates the proportion of the sex difference in ST-elevation caused by each anatomical parameter in the leads exploring the ischemic region. The larger female distance between the exploring electrode and the infarct centre was the principal contributor, accounting for 43% (anteroseptal ischemia) and 68% (inferolateral ischemia) of the reduced female ST-elevation. The larger female angle between the ECG lead axis and the infarct normal also consistently reduced female ST-elevation (19% of sex difference for anteroseptal, 14% inferolateral). Figure 6C shows ST-elevation (colour) for varying electrode position (spheres), illustrating that substantial ST-elevation was only recorded in a narrow range of electrode locations, and that female increased superior cardiac position moved the electrodes away from this range. Further details for the effect of each cardiac translation and rotation are given in Supplementary Figure 8.

Even though all anatomies had extensive, transmural ischemia affecting 16% of the myocardium, moderate translations and rotations of the heart alone were sufficient to change the diagnosis of pathology, and female anatomies were more susceptible to these changes. 75% of female anatomies changed diagnosis of ST-elevation, compared with 65% of male with current clinical guidelines. For anteroseptal ischemia, ST-elevation diagnosis was most likely to be missed with a more superior cardiac position, whereas for inferolateral ischemia, ST-elevation diagnosis was most likely to be missed with a more posterior cardiac position and high spin orientation.

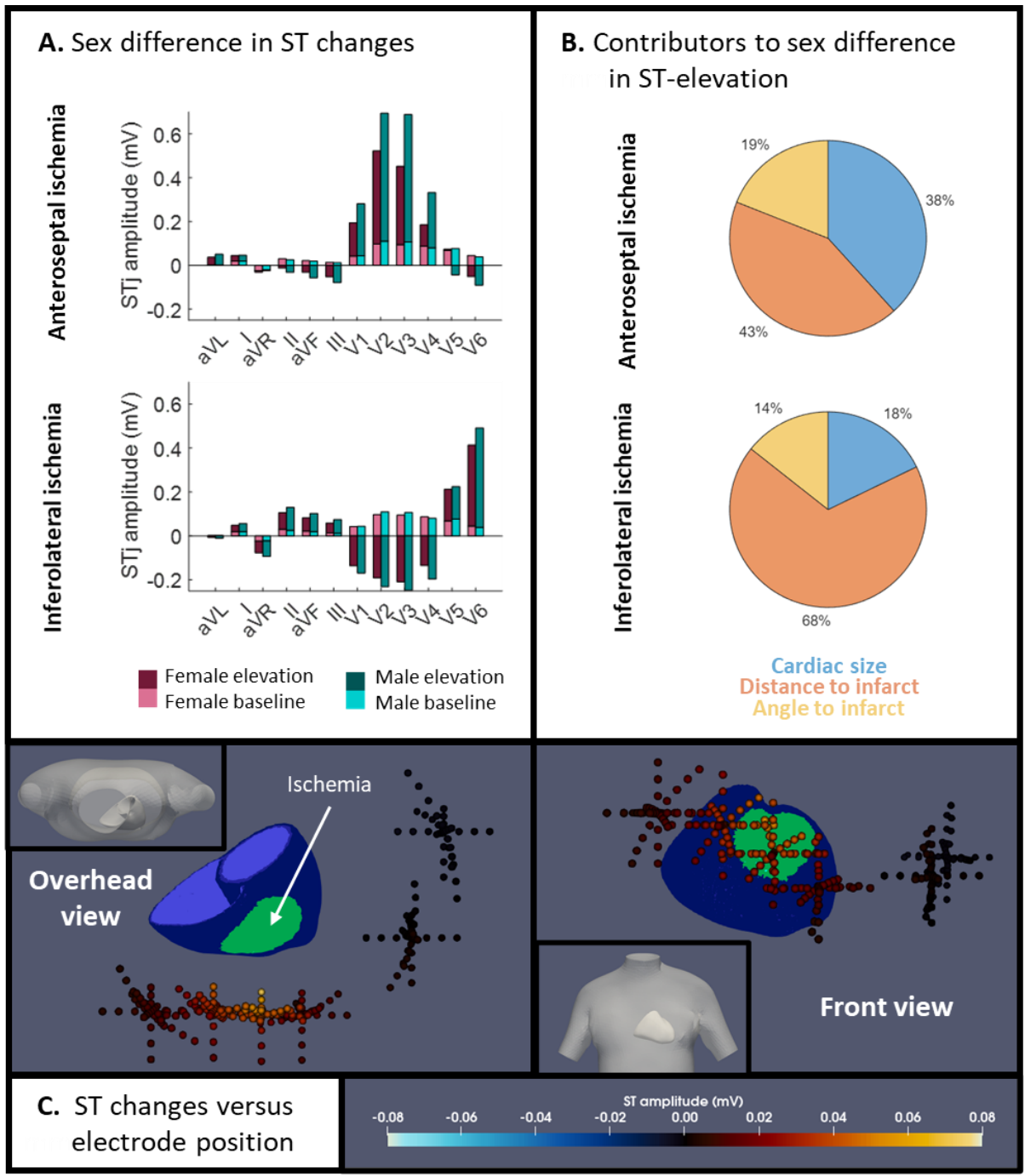

**Figure 6.** *Sex disparity of ST changes and their anatomical origins*. **A:** simulated ST amplitude for anteroseptal (upper) and inferolateral (lower) ischemia in female (pink) and male (cyan) anatomies for healthy electrophysiology (lighter) and the changes introduced by ischemia (darker). **B:** estimated proportion of sex difference in ST amplitude in leads exploring the infarct caused by cardiac size, distance from exploring electrode to infarct centre, and angle from ECG lead axis to infarct centre normal for anteroseptal (upper) and inferolateral (lower) ischemia. **C:** electrode locations (spheres) relative to the heart for a single cardiac anatomy, including rotations and translations of the heart, from an overhead (left) and front (right) view. Electrodes are coloured by their measured ST amplitude, showing that substantial ST-elevation (yellow) is only recorded in a small area near to the ischemic region (coloured green). Black electrodes recorded very little ST-elevation.

## Discussion

This study exploits the power of modelling and simulation to provide the first comprehensive computational quantification of the effect of anatomical sex differences on 8600 healthy, ischemic, and infarcted ECGs in 1720 torso-heart anatomies. Our work revealed that the ECG hallmarks of ischemia and infarction, such as interval prolongations and ST level changes were more subtle in female anatomies. This demonstrates the urgent need for reviewing clinical ECG thresholds in a more personalised manner. Credibility of simulations was established through comparison of ECG metrics and morphology in large clinical databases through infarct stages, and previous studies conducting multiscale validation studies [24, 29, 34, 42]. This combination of AI-enabled anatomical reconstruction, statistically driven hypothesis generation, and multiscale mechanistic modelling and simulation serves as a blueprint for assessing the impact of anatomical sex differences on functional markers in potentially disparate fields.

Specifically our work found that: (1) cardiac positioning alone changed diagnosis of ST-elevation in 75% of female cardiac anatomies versus 65% of male cardiac anatomies, primarily from increasing the distance from the infarct to the exploring electrode, (2) due to their smaller size, only 20% of female anatomies had QRS prolongation post-occlusion versus 55% of male, and (3) shorter female TpTe prolongation post-occlusion was caused by larger distances and angles from infarcts to ECG electrodes due to superior heart positioning inside the torso.

We found that moderate anatomical adjustment in cardiac position and orientation was sufficient to change the diagnoses of contiguous ST-elevation, even for a large area of simulated ischemia, leaving 58% of cardiac anatomies with missed diagnoses using current clinical guidelines [11]. This is consistent with experimental evidence that cardiac rotation is sufficient to make acute ischemia electrocardiographically silent [22], but goes further in including diverse anatomies and quantifying the impact of anatomical sex differences. This highlights the clinical relevance of the anatomy-ECG relationship, as ST-elevation is an important early diagnostic marker for myocardial infarction, but currently only has sex-specific thresholds in two leads (V2 and V3) [11]. The fact that female anatomies were more susceptible to missed diagnoses for both ischemic locations, even when the current sex specific thresholds in V2 and V3 were used, and were present in leads that currently have no sex specific thresholds for ST-elevation, implies that these guidelines may be inadequate in correcting for sex differences and thus may contribute to disparities in diagnosis and treatment [5, 6]. The fact that ST-elevation was more sensitive than intervals to the electrode-ischemic region distance is consistent with the biophysical understanding that electrical amplitudes reduce with the distance between the source and electrode [52]. This meant that the female more posterior and superior cardiac position substantially contributed to the reduced female ST-elevation for both ischemia locations, but also that female anatomies with smaller hearts but larger torsos were additionally reduced. This is consistent with statistical analysis of clinical ECGs [19], and suggests that BMI may be an important factor in ST-elevation threshold adjustment. Furthermore, these results highlight the particular urgency of reviewing any ECG amplitude-based clinical threshold for the effects of anatomical sex differences, beyond just those related to MI diagnosis, such as acute pericarditis and Brugada Syndrome [53, 54].

Our analysis showing that male anatomies were nearly three times more likely than female to meet criteria for moderate QRS prolongation in infarction highlights the urgent need for anatomical and demographic correction tools. Whilst QRSd is known to increase with ventricular size [19, 55, 56], previous trials to incorporate QRSd into risk stratification tools typically use fixed QRSd thresholds (e.g. 120ms for all subjects), and have had underwhelming predictive power [14, 57]. Correction for cardiac size has however already shown success in identifying candidates for cardiac resynchronisation therapy in heart failure [58]. We also showed that the larger electrode-infarct distances (relative to cardiac size) shortened female QRSd, as it led to small late depolarisations being missed. However, the strength of this relationship varied with infarction characteristics. The effect was strongest for anteroseptal ischemia/infarction. For this infarct location, the affected tissue was connected to the septum as well as the anterior wall, meaning the patch of late depolarisation was smaller and thus easier to miss. Our analysis shows that not only is demographic correction necessary, but it must be tailored to the clinical context of use. Specifically, this may mean that QRSd should be corrected differently for sex differences in diagnosing MI (e.g. identifying abnormal Q waves [11]) than in risk stratification following MI, and potentially corrected differently in each lead. As cardiac size varies with height and sex, these characteristics may be incorporated into a correction tool for clinical ECG thresholds, without the need for any image processing.

Another insight from our work is that the structure-function relationship was highly biomarker-dependent. Despite QRSd and TpTe interval both being ECG durations, cardiac position and orientation substantially impacted TpTe interval, while their effect on QRSd was more modest. This again was related to differences in their mechanistic origins. Prolonged T waves occurred due to delays in the repolarisation of infarcted and BZ tissue, secondary to potassium current reductions, which were more severe than the depolarisation delays. This meant that lead axes which recorded the long late tail along steep gradients in repolarisation had a much longer TpTe interval than those that didn't (up to ten-fold difference). This "blind spot" where substantial TpTe prolongation was not recorded occurred when the ECG lead axis was more perpendicular to these repolarisation gradients. This suggests that TpTe interval should be clinically assessed in multiple leads to avoid the blind spot. TpTe interval has been suggested as a novel post-MI risk stratification metric, either as a raw interval or corrected by normalisation with other intervals, which may partially correct for cardiac size differences [15, 51, 59]. Importantly, our work shows that each novel ECG biomarker threshold should be newly evaluated for men and women separately rather than inferring correction methods from other intervals.

This study aimed at isolating the effect of patient anatomy, without the confounding effects of other variability. Therefore, we did not choose to reproduce population variation in ECG biomarkers and morphology, which would have required additional variability in cellular electrophysiology, connectivity, activation properties, and infarct characteristics. Therefore, any adjustment to clinical parameters should take into account both computational data, and clinical ECGs from patient populations. Patients that have, or go on after their MI event to develop, conditions such as heart failure are likely to exhibit different relationships between ECG biomarkers and anatomy due to remote zone remodelling. This work also did not consider differences in electrode placement due to breast tissue, and this would likely make the heart more superior in the electrode frame for women, potentially increasing sex differences in ECG parameters. Simulated ECGs also benefit from a lack of noise or drift. This allows a very precise delineation, especially of the T wave end point. For particularly noisy ECGs, a low amplitude tail may be missed, and this could, for example, create a stronger dependence of the TpTe on the proximity to the infarct.

To conclude, this study simulated 8600 ECGs from 1720 torso-ventricular anatomies, throughout the stages of myocardial ischemia and infarction, to quantify the effect of anatomical sex differences on key ECG biomarkers. Mechanistic modelling and simulation revealed that while the electrocardiographic hallmarks of acute ischemia and infarction were more subtle for female anatomies, the origins of this discrepancy were dependent on the ECG biomarker, and infarct stage and location. Whilst prolonged ECG intervals were primarily shorter in female anatomies due to smaller cardiac anatomies, lower ST-elevation was more related to differences in cardiac position. Thus, this work both mandates and facilitates the personalisation of ECG-based diagnosis and risk stratification tools to account for demographic variation in patient anatomy.

## Acknowledgements

For the purpose of open access, the authors have applied a CC BY public copyright licence to any Author Accepted Manuscript version arising from this submission. HJS was supported by a Wellcome Trust Studentship under Grant No. 102161/Z/13/Z. AB is supported by the Royal Society University Research Fellowship (Grant No. URF\R1\221314). LR was supported by a BBSRC PhD iCASE

(BB/V509395/1) and Russell Studentship Agreement with AstraZeneca (R67719/CN001) awarded to BR. The works of AB and VG are supported by the British Heart Foundation (BHF) Project under Grant PG/20/21/35082. This work was funded by a Wellcome Trust Fellowship in Basic Biomedical Sciences to BR, supporting MH and RD (214290/Z/18/Z) and the CompBioMed 2 Centre of Excellence in Computational Biomedicine (European Commission Horizon 2020 research and innovation programme, grant agreements No. 823712) to BR and VG. MH is further supported by the Oxford BHF Centre of Research Excellence (RE/13/1/30181). This research has been conducted using the UK Biobank Resource under Application Number '40161'.

## Author Contributions

HJS, BR, AB, and VG conceptualised the study and designed the research. BR, AB, and VG supervised the research. HJS, RD, and AB constructed the cardiac and torso anatomies. HJS and LLR designed and implemented the electrophysiological simulation protocols. HJS and MH designed the ECG delineation protocol. HJS processed and analysed the data. HJS and BR prepared the original draft and reviewed and edited the manuscript. All authors discussed the results, provided comments regarding the manuscript, and agreed on the final draft.

## Data Availability

The build of MonoAlg3D and example configuration files to replicate these simulations are available here: https://github.com/hannahsmith3141/MonoAlg3D_C_HS . The meshes were constructed from UK Biobank cardiac magnetic resonance images, which can be accessed by application upon approval here: https://www.ukbiobank.ac.uk/enable-your-research/register . The cardiac meshes were reconstructed with the pipeline from Doste et al. 2025 https://doi.org/10.48550/arXiv.2503.03706 and the code is publicly here: https://github.com/rdoste/InSilicoHeartGen . The electrode locations and torso meshes were obtained with the pipeline from Smith et al. 2025 https://doi.org/10.7554/eLife.108119.1 and the code is publicly available here: https://github.com/MultiMeDIA-Oxford/TORSO-MPP . The ECG simulation delineation tool developed for this work is available here: https://github.com/MaxxHolmes/ECG_Delineation .

1. McDonough, A.A., et al., *Sex differences in renal transporters: assessment and functional consequences.* Nature Reviews Nephrology, 2024. **20**(1): p. 21–36.
2. Bhattacharjee, S., et al., *Sex difference in tDCS current mediated by changes in cortical anatomy: A study across young, middle and older adults.* Brain Stimulation, 2022. **15**(1): p. 125–140.
3. Benjamin, E.J., et al., *Heart disease and stroke statistics—2019 update: a report from the American Heart Association.* Circulation, 2019. **139**(10): p. e56–e528.
4. Rørholm Pedersen, L., et al., *Risk factors for myocardial infarction in women and men: a review of the current literature.* Current Pharmaceutical Design, 2016. **22**(25): p. 3835–3852.
5. van der Ende, M.Y., et al., *Sex-based differences in unrecognized myocardial infarction.* Journal of the American Heart Association, 2020. **9**(13): p. e015519.
6. Lawless, M., et al., *Sex differences in treatment and outcomes amongst myocardial infarction patients presenting with and without obstructive coronary arteries: a prospective multicentre study.* European Heart Journal Open, 2023. **3**(2): p. oead033.
7. Aronis, K.N., et al., *Accurate conduction velocity maps and their association with scar distribution on magnetic resonance imaging in patients with postinfarction ventricular tachycardias.* Circulation: Arrhythmia and Electrophysiology, 2020. **13**(4): p. e007792.
8. Jiang, M., et al., *Delayed rectifier K currents have reduced amplitudes and altered kinetics in myocytes from infarcted canine ventricle.* Cardiovascular Research, 2000. **48**(1): p. 34–43.
9. Solomon, S.D., et al., *Sudden death in patients with myocardial infarction and left ventricular dysfunction, heart failure, or both.* New England Journal of Medicine, 2005. **352**(25): p. 2581–2588.
10. Callans David, J. and J.K. Donahue, *Repolarization Heterogeneity in Human Post-Infarct Ventricular Tachycardia.* JACC: Clinical Electrophysiology, 2022. **8**(6): p. 713–718.
11. Thygesen, K., et al., *Fourth universal definition of myocardial infarction (2018).* European Heart Journal, 2019. **40**(3): p. 237–269.
12. Thygesen, K., et al., *Third Universal Definition of Myocardial Infarction.* Circulation, 2012. **126**(16): p. 2020–2035.
13. Moss, A.J., et al., *Prophylactic implantation of a defibrillator in patients with myocardial infarction and reduced ejection fraction.* New England Journal of Medicine, 2002. **346**(12): p. 877–883.
14. Yerra, L., et al., *Association of QRS duration and outcomes after myocardial infarction: the VALIANT trial.* Heart Rhythm, 2006. **3**(3): p. 313–316.

15. Tatlisu, M.A., et al., *Can the T-peak to T-end interval be a predictor of mortality in patients with ST-elevation myocardial infarction?* Coronary Artery Disease, 2014. **25**(5).
16. Haukilahti, M.A.E., et al., *Sudden cardiac death in women.* Circulation, 2019. **139**(8): p. 1012–1021.
17. Ilkhanoff, L., et al., *Factors associated with development of prolonged QRS duration over 20 years in healthy young adults: the Coronary Artery Risk Development in Young Adults study.* J Electrocardiol, 2012. **45**(2): p. 178–84.
18. Bolatlı, G., et al., *Examination of the levels of structures in the thorax in multidetector computerized tomography images.* Journal of Surgery and Medicine, 2020. **4**(9): p. 784–789.
19. Smith, H.J., et al., *Anatomical basis of sex differences in the electrocardiogram identified by three-dimensional torso-heart imaging reconstruction pipeline*. 2025, eLife Sciences Publications, Ltd.
20. Simone, G.d., et al., *Gender differences in left ventricular growth.* Hypertension, 1995. **26**(6): p. 979–983.
21. Rao, A.C.A., et al., *Electrocardiographic QRS duration is influenced by body mass index and sex.* IJC Heart & Vasculature, 2021. **37**: p. 100884.
22. MacLeod, R.S., R.L. Lux, and B. Taccardi, *A possible mechanism for electrocardiographically silent changes in cardiac repolarization.* Journal of Electrocardiology, 1998. **30**: p. 114–121.
23. Martinez-Navarro, H., et al., *High arrhythmic risk in antero-septal acute myocardial ischemia is explained by increased transmural reentry occurrence.* Scientific Reports, 2019. **9**(1): p. 16803.
24. Wang, Z.J., et al., *Human biventricular electromechanical simulations on the progression of electrocardiographic and mechanical abnormalities in post-myocardial infarction.* EP Europace, 2021. **23**(Supplement_1): p. i143–i152.
25. Martinez-Navarro, H., et al., *ECG analysis of ventricular fibrillation dynamics reflects ischaemic progression subject to variability in patient anatomy and electrode location.* Frontiers in Cardiovascular Medicine, 2024. **Volume 11 - 2024**.
26. Petersen, S.E., et al., *Imaging in population science: cardiovascular magnetic resonance in 100,000 participants of UK Biobank - rationale, challenges and approaches.* Journal of Cardiovascular Magnetic Resonance, 2013. **15**(1): p. 46.
27. Bai, W., et al., *Automated cardiovascular magnetic resonance image analysis with fully convolutional networks.* Journal of Cardiovascular Magnetic Resonance, 2018. **20**(1): p. 65.

28. Smith, H.J., et al. *Automated torso contour extraction from clinical cardiac MR slices for 3D torso reconstruction*. in *2022 44th Annual International Conference of the IEEE Engineering in Medicine & Biology Society (EMBC)*. 2022.

29. Sachetto Oliveira, R., et al., *Performance evaluation of GPU parallelization, space-time adaptive algorithms, and their combination for simulating cardiac electrophysiology.* International Journal for Numerical Methods in Biomedical Engineering, 2018. **34**(2): p. e2913.

30. Britton, O.J., et al., *Experimentally calibrated population of models predicts and explains intersubject variability in cardiac cellular electrophysiology.* Proceedings of the National Academy of Sciences of the United States of America, 2013. **110**(23): p. E2098–105.

31. Banerjee, A., et al., *A completely automated pipeline for 3D reconstruction of human heart from 2D cine magnetic resonance slices.* Philosophical Transactions of the Royal Society A: Mathematical, Physical and Engineering Sciences, 2021. **379**(2212): p. 20200257.

32. Bai, W., et al., *A bi-ventricular cardiac atlas built from 1000+ high resolution MR images of healthy subjects and an analysis of shape and motion.* Medical Image Analysis, 2015. **26**(1): p. 133–145.

33. Doste, R., et al., *An Automated Computational Pipeline for Generating Large-Scale Cohorts of Patient-Specific Ventricular Models in Electromechanical In Silico Trials.* arXiv preprint arXiv:2503.03706, 2025.

34. Tomek, J., et al., *Development, calibration, and validation of a novel human ventricular myocyte model in health, disease, and drug block.* eLife, 2019. **8**: p. e48890.

35. Tomek, J., A. Bueno-Orovio, and B. Rodriguez, *ToR-ORd-dynCl: an update of the ToR-ORd model of human ventricular cardiomyocyte with dynamic intracellular chloride.* bioRxiv, 2020: p. 2020.06.01.127043.

36. Doste, R., et al., *A rule-based method to model myocardial fiber orientation in cardiac biventricular geometries with outflow tracts.* International Journal for Numerical Methods in Biomedical Engineering, 2019. **35**(4): p. e3185.

37. Taggart, P., et al., *Inhomogeneous transmural conduction during early ischaemia in patients with coronary artery disease.* Journal of Molecular and Cellular Cardiology, 2000. **32**(4): p. 621–30.

38. Schuler, S., et al., *Cobiveco: Consistent biventricular coordinates for precise and intuitive description of position in the heart – with MATLAB implementation.* Medical Image Analysis, 2021. **74**: p. 102247.

39. Szentadrassy, N., et al., *Apico–basal inhomogeneity in distribution of ion channels in canine and human ventricular myocardium.* Cardiovascular Research, 2005. **65**(4): p. 851–860.
40. Boukens, B.J., et al., *Transmural APD gradient synchronizes repolarization in the human left ventricular wall.* Cardiovascular Research, 2015. **108**(1): p. 188–196.
41. Zhou, X., et al., *Clinical phenotypes in acute and chronic infarction explained through human ventricular electromechanical modelling and simulations.* eLife, 2024. **13**: p. RP93002.
42. Riebel, L.L., et al., *In silico evaluation of cell therapy in acute versus chronic infarction: role of automaticity, heterogeneity and Purkinje in human.* Scientific Reports, 2024. **14**(1): p. 21584.
43. Martinez-Navarro, H., et al., *Electrophysiological and anatomical factors determine arrhythmic risk in acute myocardial ischaemia and its modulation by sodium current availability.* Interface Focus, 2020. **11**(1).
44. Spath, N.B., et al., *Assessment of stunned and viable myocardium using manganese-enhanced MRI.* Open Heart, 2021. **8**(1).
45. Garg, P., et al., *Acute infarct extracellular volume mapping to quantify myocardial area at risk and chronic infarct size on cardiovascular magnetic resonance imaging.* Circulation: Cardiovascular Imaging, 2017. **10**(7): p. e006182.
46. Viceconti, M., et al., *In silico trials: Verification, validation and uncertainty quantification of predictive models used in the regulatory evaluation of biomedical products.* Methods, 2021. **185**: p. 120–127.
47. Goldberger, A.L., et al., *PhysioBank, PhysioToolkit, and PhysioNet: components of a new research resource for complex physiologic signals.* Circulation, 2000. **101**(23): p. e215–e220.
48. Riebel, L.L., et al. *In Silico Identification of the Key Ionic Currents Modulating Human Pluripotent Stem Cell-Derived Cardiomyocytes Towards an Adult Phenotype*. in *2021 Computing in Cardiology (CinC)*. 2021.
49. Gima, K. and Y. Rudy, *Ionic current basis of electrocardiographic waveforms.* Circulation Research, 2002. **90**(8): p. 889–896.
50. Ayachit, U., *The ParaView Guide: A Parallel Visualization Application*. 2015: Kitware, Inc.
51. Smetana, P., et al., *Sex differences in the rate dependence of the T wave descending limb.* Cardiovascular Research, 2003. **58**(3): p. 549–554.
52. van Oosterom, A., R. Hoekema, and G.J.H. Uijen, *Geometrical factors affecting the interindividual variability of the ECG and the VCG.* Journal of Electrocardiology, 2000. **33**: p. 219–227.
53. Lazarou, E., et al., *Acute Pericarditis: Update.* Current Cardiology Reports, 2022. **24**(8): p. 905–913.

54. Sieira, J. and P. Brugada, *The definition of the Brugada syndrome.* European Heart Journal, 2017. **38**(40): p. 3029–3034.

55. Mincholé, A., et al., *MRI-based computational torso/biventricular multiscale models to investigate the impact of anatomical variability on the ECG QRS complex.* Frontiers in Physiology, 2019. **10**.

56. Hakacova, N., et al., *Aspects of Left Ventricular Morphology Outperform Left Ventricular Mass for Prediction of QRS Duration.* Annals of Noninvasive Electrocardiology, 2010. **15**(2): p. 124–129.

57. Buxton, A.E., et al., *QRS duration does not predict occurrence of ventricular tachyarrhythmias in patients with implanted cardioverter-defibrillators.* Journal of the American College of Cardiology, 2005. **46**(2): p. 310–316.

58. Zweerink, A., et al., *Size Matters: Normalization of QRS Duration to Left Ventricular Dimension Improves Prediction of Long-Term Cardiac Resynchronization Therapy Outcome.* Circulation: Arrhythmia and Electrophysiology, 2018. **11**(12): p. e006767.

59. Mugnai, G., et al., *Tpeak-to-Tend/QT is an independent predictor of early ventricular arrhythmias and arrhythmic death in anterior ST elevation myocardial infarction patients.* European Heart Journal. Acute Cardiovascular Care, 2016. **5**(6): p. 473–480.

*Supplementary materials for: Female anatomies disguise ECG abnormalities following myocardial infarction: an AI-enabled modelling and simulation study*

# SM1: Methods

*SM1.1 Selection of subjects from the UK Biobank*

40 subjects (ten healthy male, ten healthy female, ten post-MI male, and ten post-MI female) were selected from a population of 1051 healthy and 425 post-MI subjects in the UK Biobank [26]. Details of this larger population selection are given in Smith et al. 2025 [19]. As BMI accounts for a large source of variability in many anatomical parameters, each subpopulation was split into BMI deciles, and one subject was chosen from each decile. This choice was made manually, maximising the spread of anatomical parameters across their distributions, as illustrated in Figures 2B-M, and replicating trends between populations, as shown in Supplementary Table 1.

**Supplementary Table 1.** *Demographic, clinical, and anatomical characteristics of the UK Biobank population selected for statistical analysis and the subpopulation selected for simulation.*

| | UK Biobank extracted population | | | |
|---|---|---|---|---|
| | Healthy male (N=470) | Healthy female (N=581) | Post-MI male (N=341) | Post-MI female (N=84) |
| Age (years) | 61.3±7.6 | 61.0±7.5 | 67.3±6.3 | 67.1±6.1 |
| BMI (kgm$^{-2}$) | 26.3±3.6 | 25.7±4.4 | 28.1±4.0 | 28.2±5.7 |
| LVEF (%) | 57.6±5.7 | 61.0±5.4 | 54.8±8.1 | 59.5±6.6 |
| Systolic BP (mmHg) | 138±16 | 130±17 | 142±20 | 139±17 |
| Diastolic BP (mmHg) | 83±10 | 78±10 | 82±11 | 81±10 |
| Total cavity volume (ml) | 326±62 | 249±42 | 320±63 | 245±47 |
| Global LV thickness (mm) | 6.1±0.6 | 5.1±0.5 | 6.3±0.7 | 5.5±0.6 |
| V6-heart centre lateral displacement (mm) | 115.6±14.6 | 111.1±16.2 | 120.5±15.5 | 117.0±17.7 |
| V6-heart centre posterior displacement (mm) | 32.9±8.7 | 28.8±7.8 | 32.9±9.4 | 26.3±7.5 |
| V6-heart centre superior displacement (mm) | -5.9±17.6 | -8.1±16.2 | -10.7±16.4 | -10.0±15.6 |
| RL-heart centre lateral displacement (mm) | -173.7±15.1 | -159.9±15.3 | -179.2±15.1 | -165.8±15.9 |
| RL-heart centre posterior displacement (mm) | -50.6±14.6 | -48.3±14.3 | -52.5±16.3 | -53.7±15.3 |
| RL-heart centre superior displacement (mm) | -205.7±17.0 | -208.6±16.4 | -209.6±16.5 | -209.5±15.8 |
| Cardiac spin (°) | 41.4±7.6 | 44.0±8.4 | 43.0±8.4 | 44.8±9.9 |
| Cardiac verticality (°) | -63.4±7.9 | -59.5±9.0 | -67.7±7.6 | -65.4±10.0 |
| Cardiac tilt (°) | -21.2±7.5 | -23.6±8.4 | -19.2±7.1 | -21.5±9.1 |
| **Population selected for simulation** | | | | |
| | Healthy male (N=10) | Healthy female (N=10) | Post-MI male (N=10) | Post-MI female (N=10) |
| Age (years) | 63.0±7.2 | 62.3±9.0 | 66.4±5.6 | 68.7±6.1 |
| BMI (kgm$^{-2}$) | 26.6±4.3 | 25.8±4.9 | 28.4±4.9 | 28.2±6.1 |
| LVEF (%) | 56.6±3.1 | 61.1±4.1 | 58.1±7.7 | 59.7±3.8 |
| Systolic BP (mmHg) | 136±12 | 121±12 | 148±25 | 136±16 |
| Diastolic BP (mmHg) | 84±8 | 77±9 | 80±15 | 82±10 |
| Total cavity volume (ml) | 331±86 | 260±47 | 309±94 | 238±66 |
| Global LV thickness (mm) | 6.1±0.6 | 5.2±0.5 | 6.4±0.9 | 5.8±0.8 |
| V6-heart centre lateral displacement (mm) | 117.8±17.4 | 114.4±16.6 | 118.2±17.3 | 119.5±18.0 |
| V6-heart centre posterior displacement (mm) | 31.9±12.6 | 25.6±12.1 | 31.3±9.0 | 27.7±5.5 |
| V6-heart centre superior displacement (mm) | -2.5±18.8 | -5.1±18.2 | -7.5±19.3 | -10.6±13.1 |
| RL-heart centre lateral displacement (mm) | -179.1±17.0 | -152.8±20.1 | -176.4±13.7 | -166.1±15.6 |
| RL-heart centre posterior displacement (mm) | -55.7±16.2 | -48.9±17.6 | -52.6±13.7 | -53.8±17.8 |
| RL-heart centre superior displacement (mm) | -200.0±19.3 | -206.1±18.8 | -204.8±13.7 | -209.6±13.7 |
| Cardiac spin (°) | 41.0±8.0 | 41.3±9.9 | 40.2±9.1 | 44.4±9.8 |
| Cardiac verticality (°) | -63.4±8.5 | -56.0±10.2 | -64.6±10.8 | -66.8±9.6 |
| Cardiac tilt (°) | -17.1±8.3 | -20.9±9.0 | -14.5±10.0 | -19.0±6.7 |

Parameters are shown as means and associated standard deviations.

*SM1.2 Cardiac mesh reconstruction and manipulation*

The cardiac orientation was described using the three Euler angles between the torso and cardiac axes, as illustrated on Supplementary Figure 1. The cardiac X-Y-Z axes were first defined from the cardiac magnetic resonance imaging as follows. The X-axis was the intersection between the mid short axis slice and the long axis four-chamber plane. The Z-axis was the line on the long axis four-chamber plane that was perpendicular to the X-axis, i.e. the cardiac long axis. The Y-axis was simply the normal of the X-Z plane. The Euler angles describe a series of sequential rotations required to transform the torso axis angle to the cardiac axis angles. Here, x' denotes the projection of the cardiac X axis on the torso x-y plane. Firstly, α was the angle through which the torso x-y plane was 'spun' around to give the cardiac X-Y plane, which will be denoted "spin". This is calculated as the angle between the torso x-axis and x'. Secondly, β was the rotation around x', which affects the verticality of the cardiac long axis in the torso frame. This is given by the angle between the cardiac Z-axis and the torso z-axis. Thirdly, γ was the angle rotated around the cardiac long axis to essentially "tilt" the cardiac short axis plane. This is given by the angle between the cardiac X-axis and x'. The final hexahedral meshes had a side length of 500μm.

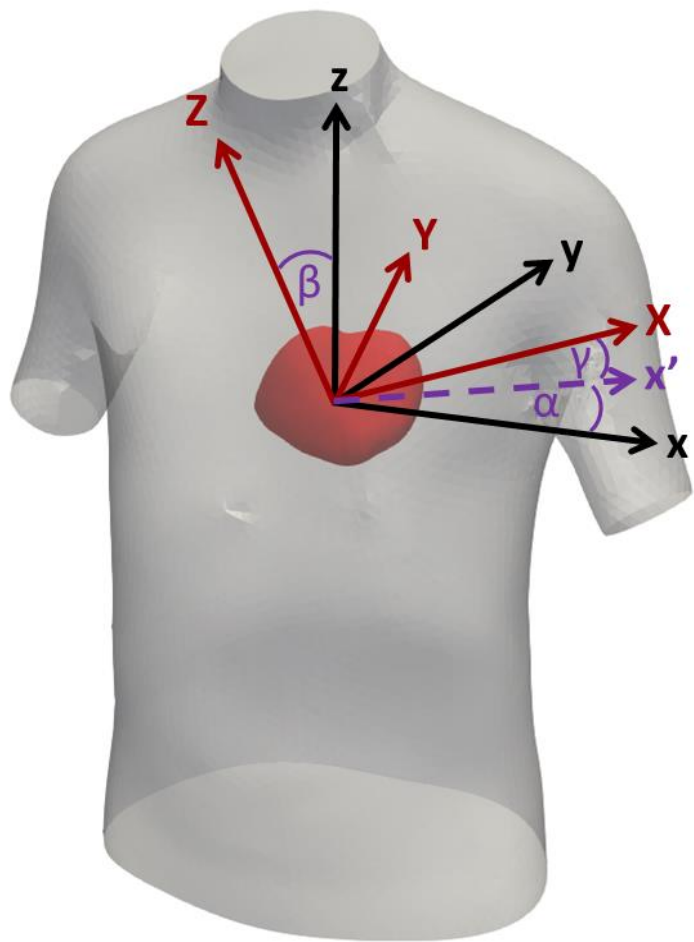

**Supplementary Figure 1.** *Cardiac-torso orientation.* Torso axes are shown in black and lower case, cardiac axes are shown in red and upper case. x' denotes the projection of the cardiac X axis onto the torso short axis (x-y) plane. α denotes the spin of the x' plane around the torso z-axis, β denotes

the verticality of the cardiac Z-axis, γ denotes the tilt of the cardiac X-Y plane around the cardiac Z-axis.

### SM1.3 Conduction velocities across infarct stages

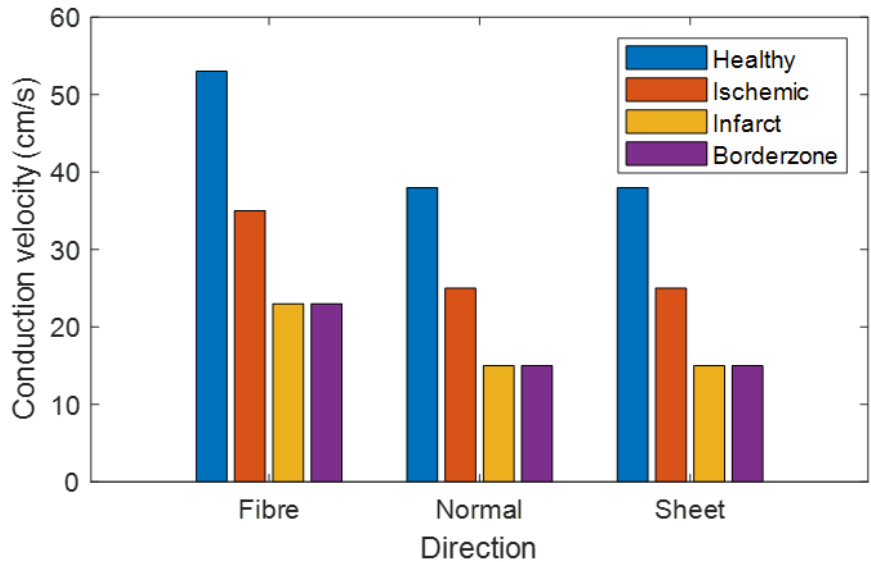

**Supplementary Figure 2.** *Achieved conduction velocities for healthy, acutely ischemic, and infarcted simulations.* Conduction velocities were tested in a 20 × 7 × 7mm slab of endocardial tissue in the fibre, sheet, and normal directions. Infarct stage is represented by bar colour for the healthy, acutely ischemic, healing stage infarct zone, and healing stage borderzone tissue.

### SM1.4 Electrophysiological remodelling in ischemia and infarction

**Supplementary Table 2.** *Remodelling parameters for cellular electrophysiology in simulations of acute ischemia and infarction.*

| Parameter | Acute ischemia | BZ | Infarct |
|---|---|---|---|
| $k_o$ | 8.5mM | 5.0mM | 5.0mM |
| IKatp | 0.07 | 0 | 0 |
| INa | 0.75 | 0.4 | 0.4 |
| ICaL | 0.75 | 0.64 | 0.64 |
| IKr | 1 | 0.7 | 0.375 |
| Ito | 1 | 0 | 0 |
| IK1 | 1 | 0.6 | 0.6 |
| ICab | 1 | 1.33 | 1.33 |
| aCaMK | 1 | 1.5 | 1.5 |
| Tau_relp | 1 | 6 | 6 |

BZ: borderzone surrounding the infarct. $k_O$: extracellular potassium concentration. All subsequent values are multipliers of specific parameters, or for channels, multipliers of the channel's maximum conductance. IKatp: ATP-sensitive potassium current, note that in the healthy and infarcted case this current is switched off. INa: fast sodium current. ICaL: long-acting calcium current. IKr: rapid delayed rectifier potassium current. Ito: transient outward potassium current. IK1: primary inward rectifier potassium current. ICab: background calcium current. aCaMK: Ca2/calmodulin-dependent protein kinase II (CaMKII) autophosphorylation rate, Tau_relp: CaMK II phosphorylation rate. Note that increases in the latter two parameters slow calcium release kinetics.

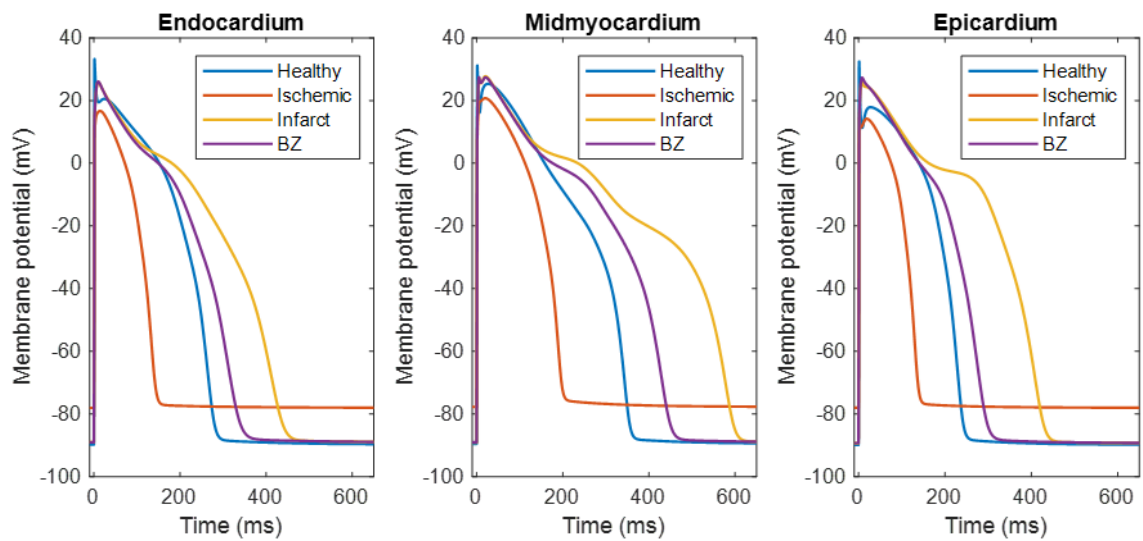

**Supplementary Figure 3.** *Effect of the ischemic and infarct remodelling on the action potential of each cell type.* Left: endocardial cell, middle: midmyocardial cell, right: epicardial cell. Infarct: infarct core zone, BZ: infarct borderzone. Acutely ischemic cells have a shorter action potential duration with a higher resting membrane potential and infarcted cells have a prolonged action potential.

*SM1.5 Uncertainty quantification*

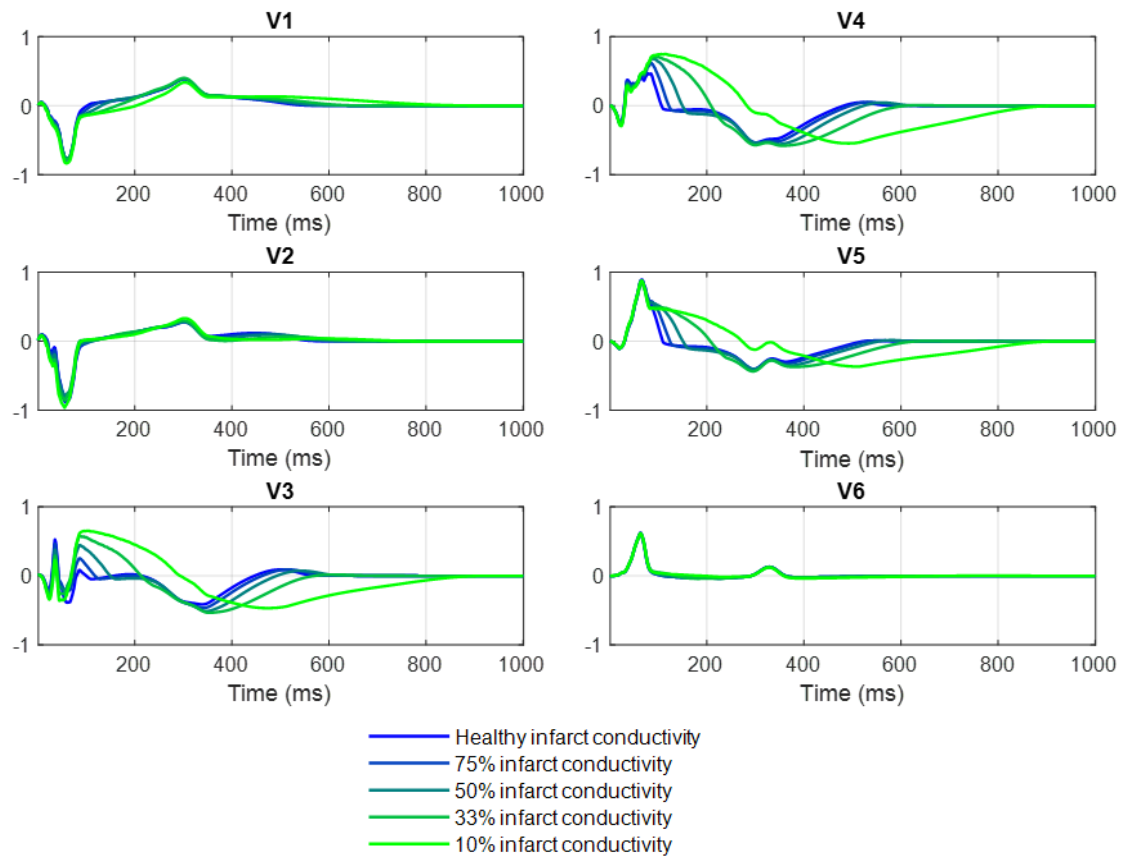

**Supplementary Figure 4.** *Sensitivity of the infarcted ECG to the conductivity of the affected region.* Precordial ECG traces for a singular test anatomy, all with infarct electrophysiology (as in Supplementary Table 2) for varying conductivity in the affected region (infarct core plus borderzone).

*SM1.6 ECG delineation*

Due to the severe ECG morphological changes in the post-MI simulations, and to exploit the lack of recording noise and drift, a custom MATLAB script was developed to delineate the ECG. Code is available in https://github.com/MaxxHolmes/ECG_Delineation . The method operates on the principle of gradient analysis of sliding windows, as follows:

(1) Define QRS onset
   Objective: QRS starts when the gradient of the ECG is changing sufficiently rapidly.
   Method: find the absolute value of the second derivative of the voltage vector with respect to time. Calculate the mean value of this in a window with a width of 3 either side of its centre.

If this is above a threshold then the first index in the window is the QRS start. The threshold, like all subsequently used, is a multiple of the maximum absolute value of the voltage, to allow for differences of voltage scales between subjects and leads. If the value is below the threshold, the centre of the sliding window moves in the positive *t* direction until the criteria has been met.

(2) *Determine QRS offset*

Objective: QRS ends when the gradient is relatively low and unchanging (though the ST segment is not guaranteed to be level or straight).

Method: for 30ms onwards following the stimulation time, find the absolute value of the second derivative of the voltage vector. Calculate the mean value of this in sliding windows with a width either side of 2 until the end of the trace. Window centres with values below a threshold were taken as candidates for the QRS offset. Then for each candidate centre, a similar sliding window of the mean absolute value of the gradient of the voltage (first derivative) was calculated with a width either side of 5. The earliest candidate centre with this value below a threshold was taken as the QRS offset.

Error flagging: The maximum activation time of each simulation was calculated as three scaled median absolute deviations above the median. If the QRS offset time was 25ms or more different from the maximum activation time, then the ECG delineation was manually checked.

(3) *Determine T peak time*

Objective: the time where the peak voltage was reached after the QRS and ST segments are finished.

Method: find the time at which the absolute value of the voltage is highest from 100ms after the QRS end to the end of the beat. T wave amplitude was defined as the value of the voltage at this point.

(4) *Determine T wave offset*

Objective: as the model does not give delayed afterdepolarisations, simply find the last point where there is a significant change in voltage.

Method: calculate the absolute value of the gradient of the voltage. Calculate the mean value of this in a sliding window, width 3, on either side, starting at the end of the beat. Move the centre of the sliding window in the negative *t* direction (earlier times) until this mean value is above a threshold, and set this as the T offset.

(5) *Calculate composite values*

QRS duration was defined as QRS offset – QRS onset. T-peak-to-end interval was defined as T offset – T peak time. ST amplitude was recorded 20ms after QRS offset to allow for any discrepancies in the QRS offset time and ensure that the QRS had fully finished.

*SM1.7 ECG amplitudes*

In order to compare simulated ECG amplitudes with clinical thresholds, they were converted to mV by multiplying by 0.0392. This value was chosen as the ratio between the peak-to-peak QRS amplitude of the healthy UKB clinical population and the corresponding healthy simulated value. In line with clinical guidelines, an ECG was considered to be ST-elevated when the ST amplitude was elevated in at least two contiguous leads. ST-elevation was defined as above 0.2mV for male anatomies, and 0.15mV for female anatomies, both in V2 and V3, and above 0.1mV for either sex otherwise [11].

*SM1.8 Statistical methods*

The anatomical parameter sets for the linear regression were checked for collinearity by calculating the variance inflation factor. This was done separately for the two infarct locations (anteroseptal and inferolateral).

**Supplementary Table 3.** *Variance inflation factor calculation for anatomical parameter sets.*

| VIF | Cardiac size | Relative distance | Angle |
|---|---|---|---|
| **Anteroseptal** | 1.1 | 1.4 | 1.3 |
| **Inferolateral** | 1.1 | 1.2 | 1.1 |

Variance inflation factor (VIF) calculation to show collinearity in linear regression parameter sets for the anteroseptal and inferolateral infarct locations. Cardiac size refers to total myocardial volume, relative distance to the distance between the exploring electrode and infarct centre divided by the cube root of the myocardial volume, and angle to the angle between the ECG lead axis and the normal to the cardiac surface at the infarct centre.

Additional figures are presented in Section SM2.3 showing the effect of changing each positional and orientational parameter on the ECG biomarkers for each lead separately. They were obtained as follows. Each cardiac anatomy was treated separately, and the median was taken at the end. The regression coefficients of the ECG biomarkers were calculated against one cardiac positional/orientational parameter with simulations whose anatomies only changed in that positional/orientational parameter, for example simulations that only varied in the cardiac inferior-superior position. Median regression coefficients over the 40 baseline anatomies were then calculated, and multiplied by the maximum variation in the anatomical parameter (e.g. maximum angular variation of 40°) for easy comparison between parameters, before being displayed as heatmaps.

# SM2: Results

*SM2.1 Sex differences in relative distance and angle to infarct*

**Supplementary Table 4.** *Relative distance between exploring electrode and infarct centre by sex.*

| Relative distance to infarct ("Cardiac lengths") | Anteroseptal | | Inferolateral | |
|---|---|---|---|---|
| Lead | Male | Female | Male | Female |
| V1 | 2.1±0.3 | 2.3±0.4 | 2.9±0.2 | 3.1±0.3 |
| V2 | 1.6±0.2 | 1.7±0.4 | 2.4±0.2 | 2.6±0.3 |
| V3 | 1.5±0.2 | 1.7±0.4 | 2.2±0.2 | 2.4±0.4 |
| V4 | 1.7±0.3 | 1.9±0.4 | 2.2±0.2 | 2.4±0.4 |
| V5 | 2.1±0.4 | 2.4±0.5 | 2.0±0.3 | 2.3±0.5 |
| V6 | 2.7±0.5 | 2.9±0.5 | 2.1±0.4 | 2.2±0.5 |

Distance between the exploring electrode for precordial leads and the centre of an anteroseptal and inferolateral infarct, measured in "cardiac lengths". This is calculated as the raw distance divided by the cube root of the myocardial volume. Values shown are mean ± standard deviation. Relative to their cardiac size, female electrodes are consistently further from the infarct, largely caused by their more superior cardiac position.

**Supplementary Table 5.** *Angle between ECG lead axis and infarct normal by sex.*

| Angle to infarct (°) | Anteroseptal | | Inferolateral | |
|---|---|---|---|---|
| Lead | Male | Female | Male | Female |
| V1 | 50.1±15.7 | 51.7±15.7 | 19.0±9.8 | 25.8±11.4 |
| V2 | 38.9±14.8 | 41.8±16.0 | 38.1±14.5 | 44.3±15.7 |
| V3 | 44.7±14.2 | 49.5±12.1 | 63.7±15.4 | 67.5±14.8 |
| V4 | 59.5±14.2 | 63.9±10.9 | 77.8±7.9 | 73.4±9.5 |
| V5 | 76.7±7.4 | 79.0±10.7 | 48.1±12.0 | 49.1±15.8 |
| V6 | 74.3±12.7 | 73.5±9.4 | 25.0±11.7 | 28.2±8.0 |

Angle between the ECG lead axis and the normal to the cardiac surface at the centre of an anteroseptal and inferolateral infarct. Values shown are mean ± standard deviation. Female anatomies have ECG lead axis consistently at a larger angle to the infarct normal, largely caused by their more superior cardiac position, alongside orientation differences in the cardiac axes.

*SM2.2 Sex differences in ECG parameter changes in ischemia and infarct*

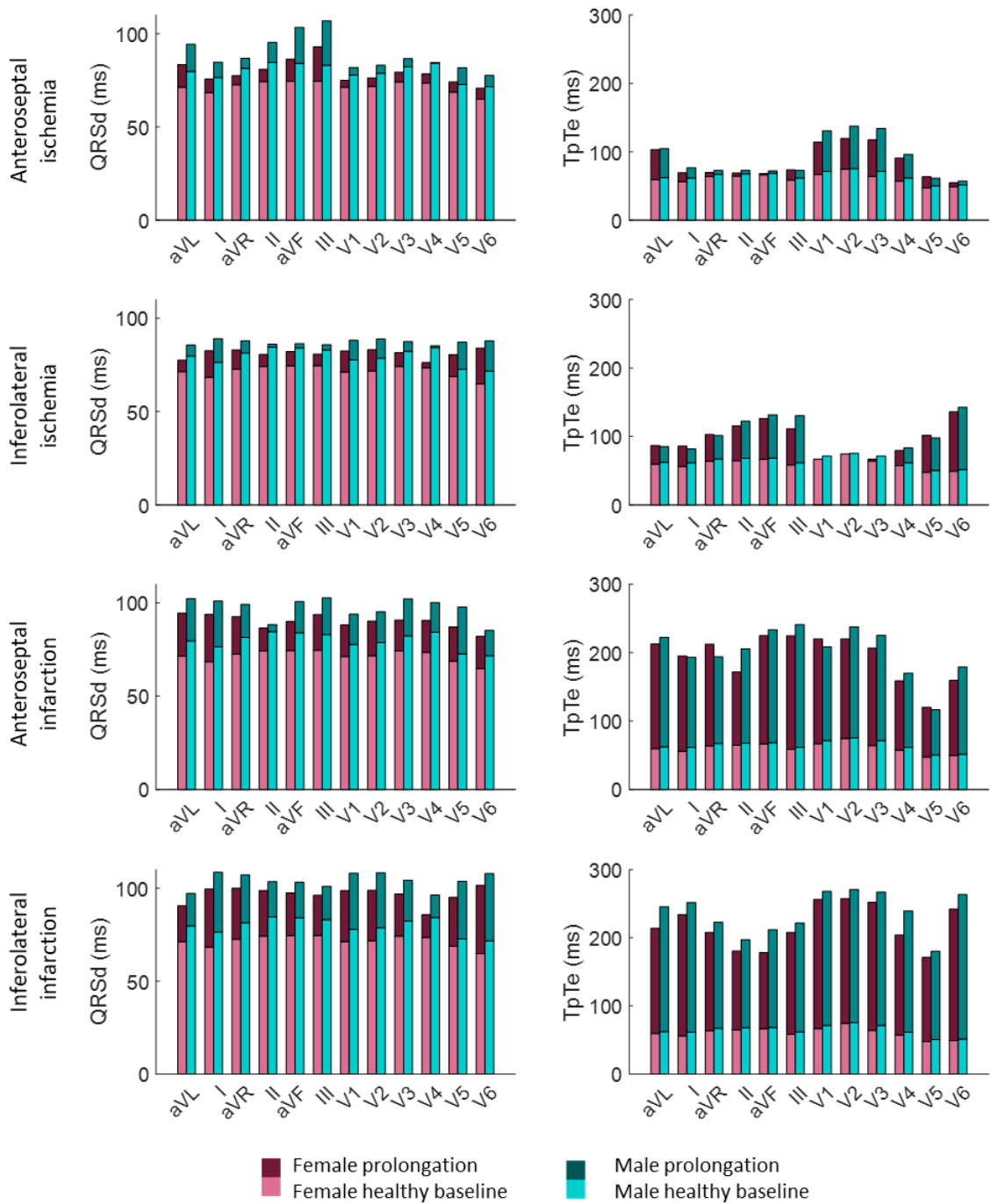

**Supplementary Figure 5.** *Sex differences in interval prolongations following occlusion across leads.* Cyan bars represent male anatomies, pink female, lighter is healthy and darker is the difference introduced by the occlusion. Values are the means across subjects. Upper to lower: anteroseptal acute ischemia, inferolateral acute ischemia, anteroseptal infarction, inferolateral infarction, left: QRS duration (QRSd), right: T-peak-to-end interval (TpTe).

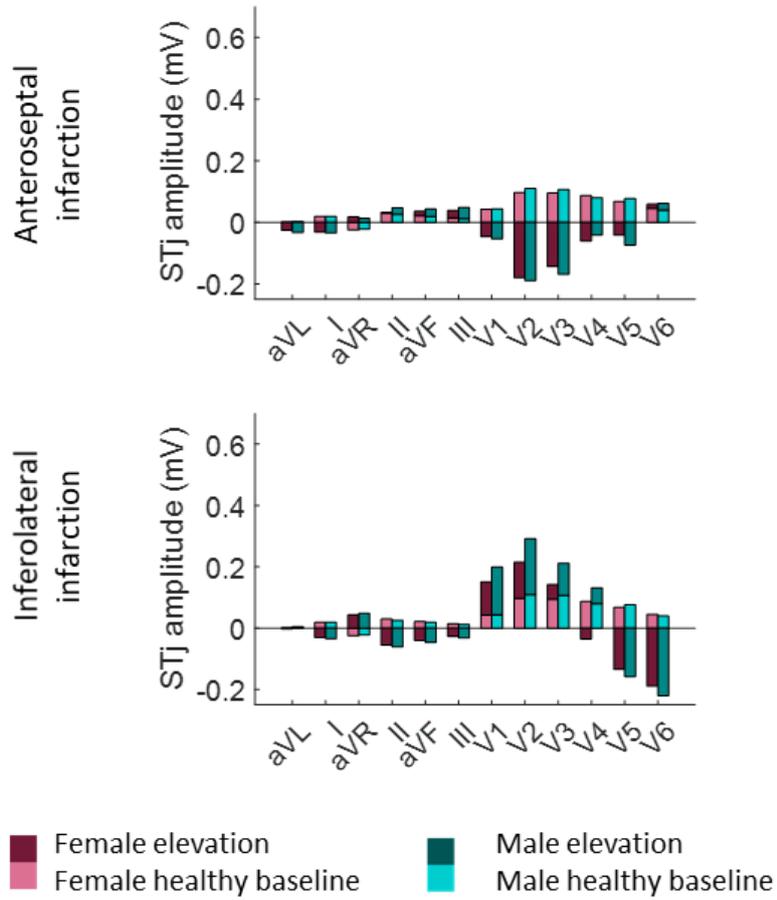

**Supplementary Figure 6.** *Sex differences in STj amplitude changes in infarction across leads.* Cyan bars represent male anatomies, pink female, lighter is healthy and darker is the difference introduced by anteroseptal (upper) and inferolateral (lower) infarction. Values are the means across subjects.

*SM2.3 Dependence of ECG biomarkers on cardiac position and orientation*

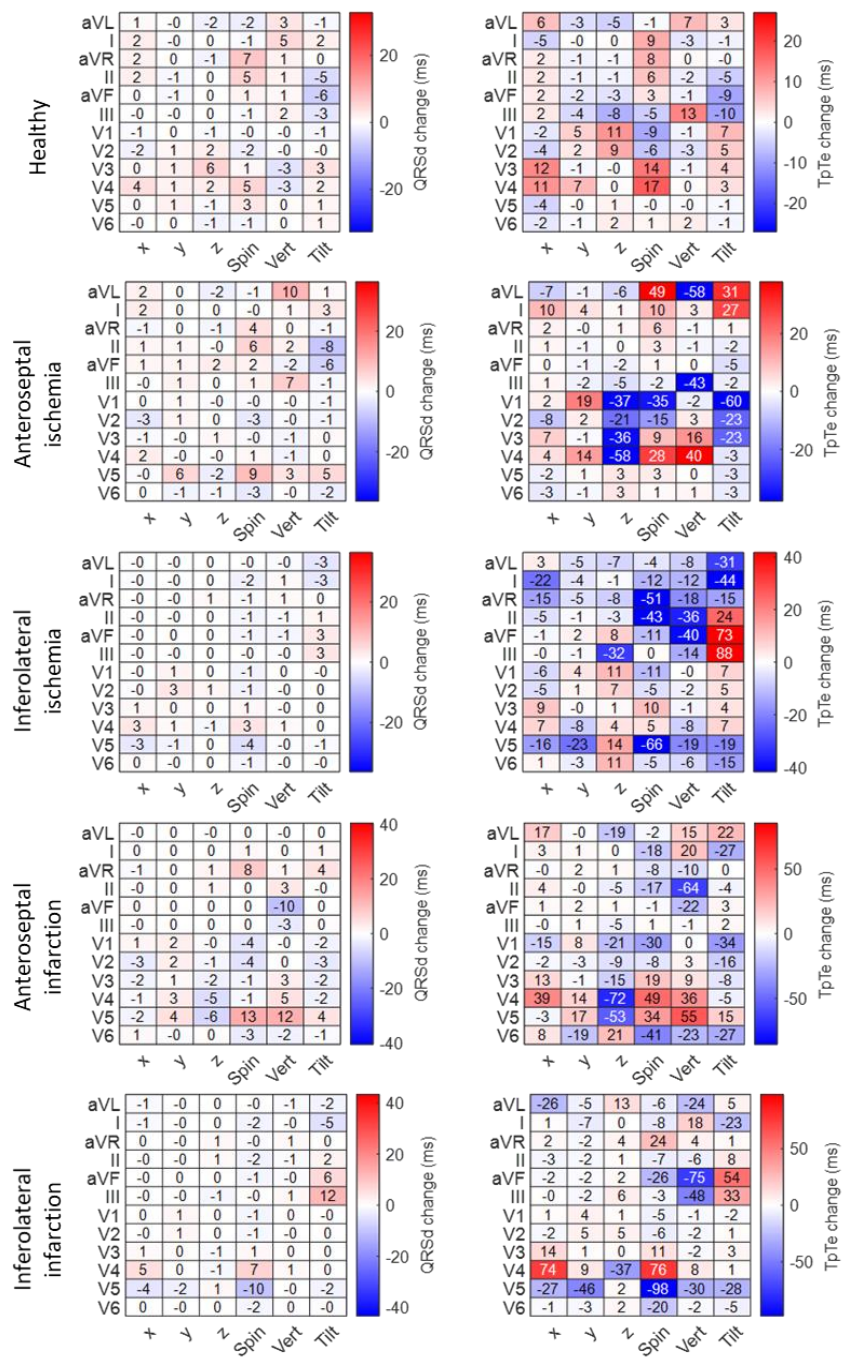

**Supplementary Figure 7.** *Variation of QRS duration (QRSd) and T-peak-to-end (TpTe) interval with cardiac affine transformations for healthy, acutely ischemic, and infarcted simulations.* Left: heatmaps of median regression coefficient of QRSd versus cardiac position and orientation across all subjects multiplied by the range of the anatomical parameter (e.g. estimating the QRSd change from moving cardiac centre 3cm medially to 3cm laterally) for upper to lower: healthy, anteroseptal acute ischemia, inferolateral acute ischemia, anteroseptal infarction, inferolateral infarction. Transformations include translations in the x (medial-lateral), y (anterior-posterior) and z (inferior-

superior) directions and cardiac rotations to spin around the torso long axis, in the horizontal-vertical direction (vert) and around the cardiac long axis to tilt the cardiac short axis plane. Right: corresponding heatmaps for TpTe against anatomical parameters. Scales have been adjusted by the mean value of the parameter across leads and subjects for ease of comparison.

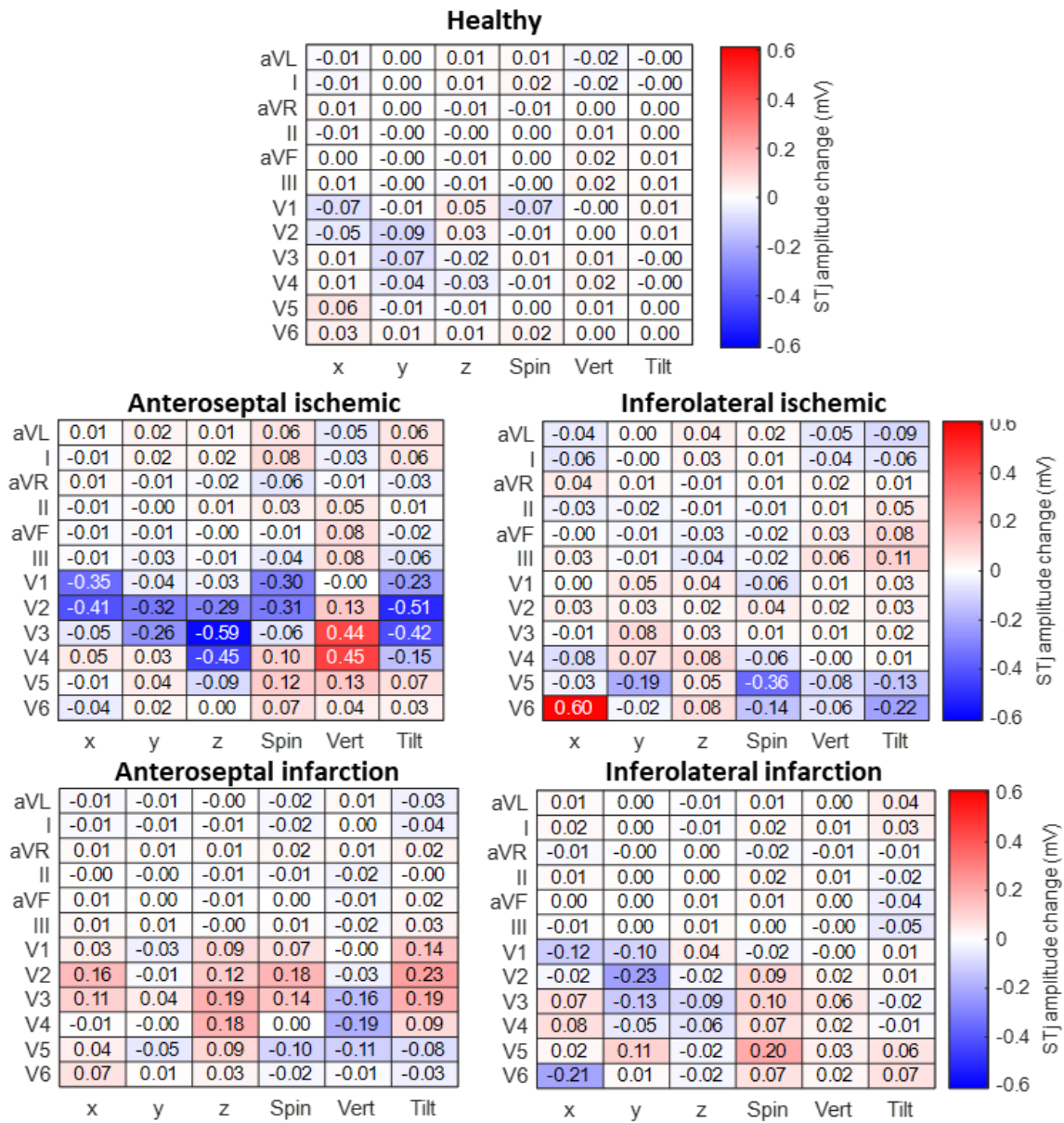

**Supplementary Figure 8.** *Variation of STj amplitude with cardiac affine transformations for healthy, acutely ischemic and infarcted simulations.* Heatmaps of median regression coefficient of STj amplitude versus cardiac position and orientation across all subjects multiplied by the range of the anatomical parameter (e.g. estimating the STj amplitude change from moving cardiac centre 3cm medially to 3cm laterally) for healthy, anteroseptal acute ischemia, inferolateral acute ischemia,

anteroseptal infarction, inferolateral infarction. Transformations include translations in the x (medial-lateral), y (anterior-posterior) and z (inferior-superior) directions and cardiac rotations to spin around the torso long axis, in the horizontal-vertical direction (vert) and around the cardiac long axis to tilt the cardiac short axis plane.

1. Petersen, S.E., et al., *Imaging in population science: cardiovascular magnetic resonance in 100,000 participants of UK Biobank - rationale, challenges and approaches.* Journal of Cardiovascular Magnetic Resonance, 2013. **15**(1): p. 46.
2. Smith, H.J., et al., *Anatomical basis of sex differences in the electrocardiogram identified by three-dimensional torso-heart imaging reconstruction pipeline*. 2025, eLife Sciences Publications, Ltd.
3. Thygesen, K., et al., *Fourth universal definition of myocardial infarction (2018).* European Heart Journal, 2019. **40**(3): p. 237–269.